\documentclass[a4paper,11pt]{article}
\usepackage{amssymb}

\usepackage{amsmath}

\newtheorem{Kprop}{\sc \textbf{Kostant's Proposition}}

\newtheorem{prop}{\sc \textbf{Proposition}}[section]
\newtheorem{cor}{\sc \textbf{Corollary}}[section]
\newtheorem{lemma}{\sc \textbf{Lemma}}[section]

\newtheorem{Def}{\sc \textbf{Definition}}[section]
\newcommand{\EQ}{\begin{equation}}
\newcommand{\EN}{\end{equation}}
\newcommand{\bea}{\begin{eqnarray}}
\newcommand{\eea}{\end{eqnarray}}

\begin{document}



\thispagestyle{empty} \hfill\ \bigskip\ \bigskip\ \bigskip\ \bigskip\
\bigskip\ 

\begin{center}
{\Large A new Rational Conformal Field Theory \\[0pt]
extension of the fully degenerate W}$_{\text{1+}\infty }^{\text{(m)}}$

{\Large \vspace{8mm}}\ \bigskip\ \bigskip

{\large Gerardo Cristofano\footnote{{\large {\footnotesize Dipartimento di
Scienze Fisiche,}\textit{\ {\footnotesize Universit\'{a} di Napoli
``Federico II''\ and INFN, Sezione di Napoli}-}\newline
{\small Via Cintia - Compl. universitario M. Sant'Angelo - 80126 Napoli,
Italy}}}, Vincenzo Marotta}$^{1}${\large , }

{\large Giuliano Niccoli\footnote{{\large {\footnotesize Laboratoire de
physique,} {\footnotesize Ecole\ Normale\ Sup\'{e}rieure\ de\ Lyon\ 46,\
all\'{e}e\ d'Italie\ 69364\ Lyon\ cedex\ 07,\ France.}\  }
\par
{\large {\footnotesize E-mail: }{\footnotesize Giuliano.Niccoli@ens-lyon.fr}}%
} \vspace{1cm}}\ \bigskip\ \bigskip

\textbf{Abstract\\[0pt]
}
\end{center}

We found new identities among the Dedekind $\eta $-function, the characters
of the $\mathcal{W}_{m}$ algebra and those of the level 1 affine Lie algebra 
$\widehat{su(m)}_{1}$. They allow to characterize the $\mathbb{Z}_{m}$%
-orbifold of the $m$-component free bosons $\widehat{u(1)}_{\mathbf{K}%
_{m,p}} $ (our theory TM) as an extension of the fully degenerate
representations of $W_{1+\infty }^{(m)}$. In particular, TM is proven to be
a $\Gamma _{\theta } $-RCFT extension of the chiral fully degenerate $%
W_{1+\infty }^{(m)}$.

\begin{quotation}
\vspace{0.3in}\ \bigskip\ \bigskip

{\footnotesize PACS: 11.25.Hf, 71.10.Pm, 73.43.Cd}

{\footnotesize Keyword: Vertex operator, Rational\ Conformal Field Theory,
Orbifold, W-algebra}
\end{quotation}

\newpage \baselineskip=18pt \setcounter{page}{2}

\section{Introduction}

In this paper, we show that our theory TM, introduced in \cite{Jain} to
describe a quantum Hall fluid at Jain fillings, gives a new rational
conformal field theory (RCFT) extension of the $W_{1+\infty }^{(m)}$ chiral
algebra corresponding to the so called irreducible fully degenerate
representations \cite{KR,FKRW,AFMO}.

A chiral algebra $\mathfrak{A}$\ in a CFT is generated by the modes (the
Fourier components) of the conserved currents; the Virasoro algebra \cite
{Virasoro,BPZ} is the chiral algebra corresponding to the analytic component 
$T(z)$ of the stress-energy tensor. A CFT can be characterized by the
corresponding chiral algebra and the set of its irreducible positive energy
(highest weight) representations closed under the fusion algebra. A
mathematical introduction to the subject of vertex or chiral algebras can be
found in \cite{Kac-b,K-T} and references there in.

An extended chiral algebra $\mathfrak{A}^{Ex.}$ is itself a chiral algebra
obtained by adding to the original $\mathfrak{A}$ the modes of further
conserved currents. The highest weight (h.w.) representations of $%
\mathfrak{A}^{Ex.}$ are opportune collections of h.w. representations of $%
\mathfrak{A}$, so that any h.w. $\mathfrak{A}^{Ex.}$-module is the direct
sum of the corresponding collection of h.w. $\mathfrak{A}$-modules.

Let us remember that RCFTs are CFTs with a finite set of h.w.
representations closed under the fusion algebra \cite{Verlinde}. The
representations of the Virasoro algebra with central charge $c\geq 1$ are
not RCFTs\ \cite{Cardy}, and so RCFTs with $c\geq 1$ correspond always to
extensions of the Virasoro algebra. An RCFT can be defined \cite{DV3} by
reorganizing the set (possibly infinite) of Virasoro h.w. representations of
the CFT into a finite number of their collections closed under the fusion
algebra, the last ones being the h.w. representations of the RCFT.

Let $\mathcal{X}$ be the finite set parametrizing the h.w. representations
of the RCFT, then the fusion algebra is defined on $A(\mathcal{X}%
):=\bigoplus_{x\in \mathcal{X}}\mathbb{C}x$ by introducing the product of
representations: 
\begin{equation}
x\circ y:=\sum_{z\in \mathcal{X}}N_{x,y,z}Cz\text{,}
\end{equation}
where $N_{x,y,z}$ are called fusion coefficients and $C$ is the finite
dimensional matrix representing the charge conjugation. The Verlinde formula
for an RCFT \cite{Verlinde,MS} expresses the fusion coefficients $N_{x,y,z}$
as a function of the elements of the symmetric unitary finite dimensional
matrix $S=\left\| S_{x,y}\right\| _{x,y\in \mathcal{X}}$, representing the
action of the modular transformation $S\in PSL(2,\mathbb{Z}):=SL(2,\mathbb{Z}%
)/\mathbb{Z}_{2}$ on the characters of the RCFT, or explicitly: 
\begin{equation}
N_{x,y,z}=\sum_{a\in \mathcal{X}}S_{a,x}S_{a,y}S_{a,z}/S_{a,e}\text{,}
\label{Verlinde-formula}
\end{equation}
where $e\in \mathcal{X}$ parametrizes the unique h.w. representation
including the vacuum vector. Furthermore, the action of the charge
conjugation on the h.w. representations is given by $C=\left\|
S_{x,y}\right\| _{x,y\in \mathcal{X}}^{2}$. The fusion algebra $(A(\mathcal{X%
}),\circ )$ is then a finite dimensional commutative associative semisimple
algebra with unity $e$.

The Verlinde formula $\left( \ref{Verlinde-formula}\right) $, in particular,
makes possible to characterize an RCFT by its properties under modular
transformations.

Let $\tilde{\Gamma}$ be a subgroup of the modular group $PSL(2,\mathbb{Z})$
containing $S$, then a CFT whose characters define a finite dimensional
representation of $\tilde{\Gamma}$ is an RCFT. In the following, we denote
this kind of RCFT as a $\tilde{\Gamma}$-RCFT to underline the subgroup $%
\tilde{\Gamma}$ of the modular group $PSL(2,\mathbb{Z})$.

These concepts are here applied to define a new RCFT extension of the CFT
with chiral algebra $W_{1+\infty }^{(m)}$ and with h.w. representations the
irreducible fully degenerate ones \cite{KR,FKRW,AFMO}. From now on, we
simply refer to such a CFT as to the fully degenerate $W_{1+\infty }^{(m)}$;
as it is well known, this CFT is not a rational one. Indeed, there are
infinitely many irreducible fully degenerate representations and all are
required to be closed under fusion, as the study of the corresponding
characters and modular transformations shows \cite{FKRW}.

In the literature there are many classes of RCFT extensions of the fully
degenerate $W_{1+\infty }^{(m)}$. Examples are the affine level 1 chiral
algebra $\mathfrak{A}_{1}(u(m))$ or $\mathfrak{A}_{1}(so(2m))$, where $%
W_{1+\infty }^{(m)}$ coincides with the corresponding $U(m)$-invariant
subalgebra. More general RCFT extensions of the fully degenerate $%
W_{1+\infty }^{(m)}$ are the lattice chiral algebras $\mathfrak{A}(Q)$
associated with the compact group $U(m)$, where $Q$ is a rank $m$ integral
lattice including as a sublattice the rank $m-1$ $su(m)$ lattice (see
section 5 of \cite{K-T}). The $m$-component free bosons $\widehat{u(1)}_{%
\mathbf{K}_{m,p}}$ (described in section 4 of this paper) can be seen as a
class of examples of these RCFT extensions.

The orbifold construction is a way to define new RCFTs starting from a given
RCFT by quotienting it with a generic discrete symmetry group $G$. More
precisely, let $G$ be a discrete group of automorphisms of the chiral
algebra $\mathfrak{A}$ of the original RCFT, then the corresponding orbifold
chiral algebra $\mathfrak{A}^{G}:=\mathfrak{A}/G$ is the subalgebra of $%
\mathfrak{A}$ defined as the invariant part of $\mathfrak{A}$ under $G$. The 
$G$-orbifold RCFT with chiral algebra $\mathfrak{A}^{G}$ has a finite set of
irreducible representations that splits in two sectors. The untwisted sector
of $\mathfrak{A}^{G}$ has irreducible representations that coincide with
those of the original chiral algebra $\mathfrak{A}$ or with opportune
restrictions of them. The twisted sector of $\mathfrak{A}^{G}$ has instead
irreducible representations that cannot be expressed in terms of those of $%
\mathfrak{A}$.

In \cite{K-T}, the orbifold construction is shown to be a tool to obtain a
class of RCFT extensions of the fully degenerate $W_{1+\infty }^{(m)}$. The
lattice chiral algebras\footnote{%
Where $\mathbb{Z}^{m}$ is the rank $m$ orthonormal lattice.} $\mathfrak{A}(%
\mathbb{Z}^{m})$ is one of the above RCFT extensions of the fully degenerate 
$W_{1+\infty }^{(m)}$ with a unique irreducible representation \cite{FKRW}.
To such an RCFT\footnote{%
See Theorem 5.2 of \cite{K-T}.} is applied the orbifold construction with
respect to $G$, a discrete group of inner automorphisms\footnote{%
In particular, $G$ is a finite subgroup of $U(m)$.} of $\mathfrak{A}(\mathbb{%
Z}^{m})$, obtaining a class of RCFT extensions of the fully degenerate $%
W_{1+\infty }^{(m)}$.

In this paper, we show that our theory TM \cite{Jain}, characterized as the
cyclic permutation orbifold \cite{K-S,F-K-S} with respect to the outer
automorphisms \cite{F-S-S,F-R-S} $\mathbb{Z}_{m}$ of the chiral algebra $%
\mathfrak{A}(\widehat{u(1)}_{\mathbf{K}_{m,p}})$, is a $\Gamma _{\theta }$%
-RCFT extension of the fully degenerate $W_{1+\infty }^{(m)}$.

The results given here contain, in particular, a generalization to any prime 
$m$ of the $m=2$ special case presented in \cite{Jain}. For $m$ not a prime
number the results still hold and will be the subject of a forthcoming paper.%
\vspace{0.1in}

The paper is organized as follows. In \textbf{section 2}, we review the CFT
with chiral algebra $W_{1+\infty }^{(m)}$ and the decomposition of the
affine level 1 $\widehat{su(m)}_{1}$ characters in terms of those of the $%
\mathcal{W}_{m}$ chiral algebra \cite{BS}. In \textbf{section 3}, we derive
the main identities among the $\eta $-function of Dedekind, the characters
of the $\mathcal{W}_{m}$ chiral algebra and the characters of the affine
level 1 $\widehat{su(m)}_{1}$, evaluated at the so called principal element
of type $\rho $ of B. Kostant \cite{Kostant}. In \textbf{section 4}, we
recall the definition of the $\Gamma _{\theta }$-RCFT $m$-component free
bosons $\widehat{u(1)}_{\mathbf{K}_{m,p}}$ \cite{FZ,WZ,DM,Cappelli}. The
corresponding chiral algebra $\mathfrak{A}(\widehat{u(1)}_{\mathbf{K}%
_{m,p}}) $ is identified together with the finite set of h.w.
representations (modules). The corresponding characters and modular
transformations are given. In \textbf{section 5}, we derive our theory TM by
making the explicit $\mathbb{Z}_{m}$ cyclic permutation orbifold
construction of the $m$-component free bosons $\widehat{u(1)}_{\mathbf{K}%
_{m,p}}$. In particular, a finite set of irreducible (h.w.) representations
(modules) of the orbifold chiral algebra $\mathfrak{A}_{\text{TM}}:=%
\mathfrak{A}^{\mathbb{Z}_{m}}(\widehat{u(1)}_{\mathbf{K}_{m,p}})$ is found.
By explicitly performing the modular transformations of the corresponding
characters, we prove that they provide a unitary finite dimensional
representation of the modular subgroup $\Gamma _{\theta }$, i.e. TM is a $%
\Gamma _{\theta }$-RCFT. In \textbf{section 6, }we show, using the
identities derived in section 3, that TM gives a $\Gamma _{\theta }$-RCFT
extension of the fully degenerate $W_{1+\infty }^{(m)}$. In \textbf{section 7%
}, some final remarks are contained. Finally, we report in two appendices
some useful definitions and results. In \textbf{appendix A}, we recall the
definition of the $\Gamma _{\theta }$ subgroup of the modular group $PSL(2,%
\mathbb{Z})$. In \textbf{appendix B}, we recall the definition of the $%
\Gamma _{\theta }$-RCFT $\widehat{u(1)}_{q}$ \cite{Cappelli}, where $q$ is
odd. The corresponding chiral algebra $\mathfrak{A}(\widehat{u(1)}_{q})$ is
identified together with the finite set of the h.w. representations
(modules). The corresponding characters and their modular transformations
are also given.

\setcounter{equation}{0}

\section{The $W_{1+\infty }^{(m)}$ chiral algebra}

$W_{1+\infty }$ is the unique nontrivial central extension \cite{KP1,PRS} of
the Lie algebra $w_{\infty }$ \cite{Bakas} of the area-preserving
diffeomorphisms on the circle; its representation theory was developed in 
\cite{KR,FKRW,AFMO}. $W_{1+\infty }$ has an infinite number of generators $%
W_{m}^{\nu }$, with $\nu $ a non negative integer and $m\in \mathbb{Z}$,
satisfying the commutation relations: 
\begin{equation}
\left[ W_{n}^{\nu },W_{n^{\prime }}^{\nu ^{\prime }}\right] =(\nu ^{\prime
}n-\nu n^{\prime })W_{n+n^{\prime }}^{\nu +\nu ^{\prime }-1}+...+c\hspace{%
0.05in}\frac{\left( \nu !\right) ^{4}}{\left( 2\nu \right) !}\left( 
\begin{array}{c}
n+\nu \\ 
n-\nu -1
\end{array}
\right) \delta _{\nu ,\nu ^{\prime }}\delta _{n+n^{\prime },0}\text{,}
\end{equation}
where dots denote a finite number of similar terms involving the operators $%
W_{n+n^{\prime }}^{\nu +\nu ^{\prime }-1-2l}$. The generators $W_{n}^{\nu }$
of $W_{1+\infty }$ define the modes of a Heisenberg algebra $\widehat{u(1)}$%
, for $\nu =0$, and those of a Virasoro algebra, for $\nu =1$, with central
charge $c$. The unitary representations of $W_{1+\infty }$ have positive
integer central charge $c=m\in \mathbb{N}$ and their h.w. representations
are defined by the h.w. vectors $\left| \mathbf{r}\right\rangle $, where $%
\mathbf{r:}=(r_{1},..,r_{m})$ is an $m$-dimensional vector with real values.
The h.w. vector $\left| \mathbf{r}\right\rangle $ is defined by: 
\begin{equation}
W_{0}^{\nu }\left| \mathbf{r}\right\rangle =w_{\nu }(\mathbf{r})\left| 
\mathbf{r}\right\rangle \ \ \ \ \ for\ \text{\ }\nu \geq 0,\ \ \ \ \ \ \ \ \
\ W_{m}^{\nu }\left| \mathbf{r}\right\rangle =0\hspace{0.15in}for\ \text{\ }%
\nu \geq 0\hspace{0.1in}m>0\text{,}
\end{equation}
with eigenvalues \cite{K-T}: 
\begin{equation}
w_{\nu }(\mathbf{r})=\frac{\left( \nu -1\right) !\nu !}{\left( 2\nu \right) !%
}\sum_{j=0}^{\nu -1}\left( 
\begin{array}{c}
\nu \\ 
j
\end{array}
\right) \left( 
\begin{array}{c}
\nu \\ 
j+1
\end{array}
\right) \sum_{i=1}^{m}r_{i}(r_{i}-j)\cdots (r_{i}+\nu -j-1)\text{.}
\end{equation}
Thus, in particular, $\left| \mathbf{r}\right\rangle $ is a h.w. vector\ for
the Virasoro algebra defined by $L_{m}:=W_{m}^{1}$,$\,\,m\in \mathbb{Z}$,
with conformal dimension $h_{\mathbf{r}}:=w_{1}(\mathbf{r})=\left(
1/2\right) \sum_{i=1}^{m}r_{i}^{2}$. The unitary irreducible\ (h.w.) module $%
\mathbb{W}_{\mathbf{r}}^{(m)}$ of $W_{1+\infty }^{(m)}$ with central charge $%
c=m$ is built by the action of the generators $W_{n}^{\nu }$ on the h.w.
vector $\left| \mathbf{r}\right\rangle $ quotient the submodule generated by
the unique singular vector of degree $m+1$ \cite{KR}. The unitary
irreducible\ h.w. representations of $W_{1+\infty }^{(m)}$ are given in
terms of those of the $m$-component free bosons $\widehat{u(1)}^{\otimes m}$%
. They can be of two types: generic or degenerate. The h.w. representations
defined by $\left| \mathbf{r}\right\rangle $ are generic if $\mathbf{r}%
=(r_{1},..,r_{m})$ satisfies the conditions $r_{a}-r_{b}\notin \mathbb{Z}$, $%
\forall a\neq b\in \{1,..,m\}$, while they are degenerate if $r_{a}-r_{b}\in 
\mathbb{Z}$, for some $a\neq b$. Finally, the h.w. representations are fully
degenerate if $\mathbf{r}=(r_{1},..,r_{m})$ satisfies the conditions $%
r_{a}-r_{b}\in \mathbb{Z}$, $\forall a\neq b\in \{1,..,m\}$.

The irreducible fully degenerate representations of $W_{1+\infty }^{(m)}$
are isomorphic \cite{Cappelli} to those of $\widehat{u(1)}\bigotimes 
\mathcal{W}_{m}$, where $\mathcal{W}_{m}$ is the algebra with central charge 
$c=(m-1)$ defined by the limit $a\rightarrow \infty $ of the
Zamolodchikov-Fateev-Lukyanov algebra with $c=(m-1)\left( 1-\frac{m(m+1)}{%
a(a+1)}\right) $ \cite{ZF}. The irreducible fully degenerate representations
of $W_{1+\infty }^{(m)}$ are classified with h.w. $\mathbf{r}$ satisfying
the additional condition that its elements are arranged in a decreasing
order, that is $\mathbf{r\in }\mathbb{P}^{(m)}$, where $\mathbb{P}%
^{(m)}:=\left\{ \mathbf{r\in }\mathbb{R}^{m}:r_{1}\geq \cdots \geq r_{m},%
\hspace{0.05in}r_{a}-r_{b}\in \mathbb{Z},\hspace{0.05in}\forall a\neq b\in
\{1,..,m\}\vspace{0.08in}\right\} $. It is worth noticing that for any h.w. $%
\mathbf{r\in }\mathbb{P}^{(m)}$, defining an irreducible\ fully degenerate
representation of $W_{1+\infty }^{(m)}$, a h.w. $\Lambda $ of $su(m)$ is
defined in the following way: 
\begin{equation}
\Lambda :=\sum_{i=1}^{m-1}\lambda _{i}\Lambda _{i},\hspace{0.1in}\lambda
_{i}:=r_{i}-r_{i+1}\in \mathbb{Z}_{+}\text{,}  \label{def-Lambda}
\end{equation}
where $\Lambda _{i}$ are the fundamental weights of $su(m)$ and $\lambda
_{i} $ are the Dynkin labels.

The $\mathcal{W}_{m}$ chiral algebra can be also defined by a coset
construction of the kind $W\left[ \text{\^{g}}_{k}/\text{g};k\right] $ based
on the Casimir operators of a finite algebra g (see \cite{BS} for details).
In particular, the coset that defines $\mathcal{W}_{m}$ is $W[\widehat{su(m)}%
_{k}/su(m);k=1]$ and involves the finite algebra $su(m)$; thus, the central
charge of the CFT with chiral algebra $\mathcal{W}_{m}$ has the same value
of that of the level 1 affine $\widehat{su(m)}_{1}$, i.e. $c_{\mathcal{W}%
_{m}}=c_{\widehat{su(m)}_{1}}=m-1$.

In the following, we will make use of characters for clarifying the
relations among the representations of the chiral algebras under study.
Indeed, to any h.w. representation of a chiral algebra we can associate a
character which accounts for the main properties of the representation. The
explicit form of the character depends on the nature of the \textit{chiral
algebra}. In the particular case of a Lie algebra or of a Kac-Moody algebra
(see chapter 9 of \cite{Kac-book}) we can define the \textit{formal character%
}, corresponding to a given h.w. $\Lambda $, as the \textit{formal }%
function: 
\begin{equation}
\chi _{\Lambda }^{\text{g}}:=\sum_{\Lambda ^{\prime }\in \Omega _{\Lambda
}}mult_{\Lambda }(\Lambda ^{\prime })e^{2\pi i\Lambda ^{\prime }}\text{,}
\label{ch-g}
\end{equation}
where $\Omega _{\Lambda }$ is the set of the weights in the h.w. $\Lambda $
representation of the algebra g, $mult_{\Lambda }(\Lambda ^{\prime })$ is
the multiplicity\footnote{%
That is, $mult_{\Lambda }(\Lambda ^{\prime })$ is the dimension of the
eigenspace $V_{\Lambda }^{\Lambda ^{\prime }}$ with eigenvalue $\Lambda
^{\prime }$ in the weight space decomposition of the h.w. module $V_{\Lambda
}$: 
\begin{equation*}
V_{\Lambda }=\sum_{\Lambda ^{\prime }\in \Omega _{\Lambda }}V_{\Lambda
}^{\Lambda ^{\prime }}.
\end{equation*}
} of the weight $\Lambda ^{\prime }$ and $e^{\Lambda ^{\prime }}$ denotes a
formal exponential satisfying: 
\begin{equation}
e^{\Gamma _{1}}e^{\Gamma _{2}}=e^{\Gamma _{1}+\Gamma _{2}}\text{ and }%
e^{\Gamma }(\xi )=e^{(\Gamma ,\xi )}\text{,}
\end{equation}
where $(,)$ is the bilinear form (Killing form) on g and $\xi $ is an
arbitrary element of the dual Cartan subalgebra. The action of the
exponential $e^{\Gamma }$ on $\xi $ allows to compute $\chi _{\Lambda }^{%
\text{g}}(\xi )$, the so called \textit{specialization} of the character at $%
\xi $. In the case of Lie algebras, the group character definition, given in
the representation theory of Lie groups (see \cite{FH}), is simply related
to that of the formal character (as explained in section 13.4.1 of \cite{Di
Francesco}), so from now on we will refer to it as the character associated
to a h.w. representation of the algebra. For more general chiral algebras
the definition of character can be given analogously. In this paper, in
particular, we will define and study the characters for a class of chiral
algebras which are cyclic orbifold of lattice chiral algebras (see \cite
{Kac-b,K-T} for a general definition of the characters of lattice chiral
algebras).

Let us now consider the characters of the coset $\mathcal{W}_{m}$; by
definition, the h.w. representations of $\mathcal{W}_{m}$ are defined by
decomposing those of $\widehat{su(m)}_{1}$ in terms of those of $su(m)$.
Thus, the characters of the coset $\mathcal{W}_{m}$ are the \textit{%
branching functions} constructed by decomposing the characters of $\widehat{%
su(m)}_{1}$ in terms of those of $su(m)$: 
\begin{equation}
\chi _{\widehat{\Lambda }_{l}}^{\widehat{su(m)}_{1}}(\xi |\tau
)=\sum_{\Lambda \in P_{+}\cap \Omega _{l}}b_{\widehat{\Lambda }%
_{l}}^{\Lambda }(\tau )\chi _{\Lambda }^{su(m)}(\xi )\text{,}
\label{coset-1}
\end{equation}
where $\Omega _{l}$ is the finite part of $\widehat{\Omega }_{l}$, the set
of the affine weights in the integrable h.w. representation of $\widehat{%
su(m)}_{1}$ corresponding to the fundamental weight $\widehat{\Lambda }_{l}$%
, and $P_{+}$\ is the set of the dominant weights of the Lie algebra $su(m)$%
. Therefore, the character of $\mathcal{W}_{m}$ corresponding to the h.w. $%
\Lambda $ is defined by $\chi _{\mathbf{\ }\Lambda }^{\mathcal{W}_{m}}(\tau
):=b_{\widehat{\Lambda }_{l}}^{\Lambda }(\tau )$ and explicitly given by: 
\begin{equation}
\chi _{\Lambda }^{\mathcal{W}_{m}}(\tau )=\frac{q^{\frac{\Lambda ^{2}}{2}}}{%
\eta (\tau )^{m-1}}\prod_{\alpha \epsilon \Delta _{+}}(1-q^{\left( \Lambda
+\rho {,}\alpha \right) })\text{,}
\end{equation}
where $q:=e^{2\pi i\tau }$, $\rho :=\left( \sum_{\alpha \epsilon \Delta
_{+}}\alpha \right) /2$ is the Weyl vector, $\Delta _{+}$ is the set of the
positive roots of $su(m)$, $\Lambda =\Lambda _{l}+\mathbf{\gamma }$ with $%
\Lambda _{l}$ a fundamental weight of $su(m)$ and $\gamma \in Q$, the set of
roots of $su(m)$. While, according to the \textit{Weyl character formula},
the character corresponding to the h.w. $\Lambda $ of $su(m)$ is: 
\begin{equation}
\chi _{\Lambda }^{su(m)}(\xi )=\frac{\sum_{w\epsilon W}\epsilon (w)e^{2\pi
i(w(\Lambda +\rho ){,}\xi )}}{\sum_{w\epsilon W}\epsilon (w)e^{2\pi i(w\rho {%
,}\xi )}}\text{,}  \label{su(m)-character}
\end{equation}
where $w$ is an element in the Weyl group $W$ of $su(m)$, $\epsilon
(w)=(-1)^{l(w)}\ $is the signature of the Weyl reflection $w$ and $l(w)\in
N\ $is the length\ of\ $w$.

For $z\rightarrow 0$ the character $\chi _{\Lambda }^{su(m)}(\xi =z\rho )$
goes to the dimension of the h.w. $\Lambda $ representation of the finite
algebra $su(m)$, $d_{su(m)}(\Lambda )$. Thus, the characters of $\widehat{%
su(m)}_{1}$, specialized to the weight $\hat{\xi}=\left. \left( z\rho +\tau 
\widehat{\Lambda }_{0}\right) \right| _{z=0}$, can be written in the form: 
\begin{equation}
\chi _{l}^{\widehat{su(m)}_{1}}(\tau ):=\underset{z\rightarrow 0}{\lim }%
\text{ }\chi _{\widehat{\Lambda }_{l}}^{\widehat{su(m)}_{1}}(\xi =z\rho
|\tau )=\sum_{\Lambda \in P_{+}\cap \Omega _{l}}d_{su(m)}(\Lambda )\chi
_{\Lambda }^{\mathcal{W}_{m}}(\tau )\text{.}  \label{su(m)-Wm}
\end{equation}

The above analysis on the h.w. $\Lambda $ representations of $\mathcal{W}%
_{m} $ and the characterization given of the irreducible fully degenerate
h.w. $\mathbf{r\in }\mathbb{P}^{(m)}$ of $W_{1+\infty }^{(m)}$ imply the
following expression for the corresponding character: 
\begin{equation}
\chi _{\mathbf{r}}^{\mathbf{w}_{m}}(w|\tau ):=Tr_{\mathbb{W}_{\mathbf{r}%
}^{(m)}}\left( e^{2\pi i\tau \left( L_{0}-\frac{m}{24}\right) }e^{2\pi
iwJ}\right) =\chi _{\Lambda }^{\mathcal{W}_{m}}(\tau )\frac{e^{2\pi i\left\{
\tau \frac{1}{2m}\left( \sum_{i=1}^{m}r_{i}\right) ^{2}+w\mathbf{rt}\det (%
\mathbf{R})\right\} }}{\eta \left( \tau \right) }\text{,}  \label{ch-w-m}
\end{equation}
where $\mathbf{t^{T}:=}\left( 1,..,1\right) $, $L_{0}$ is the zero mode of
the Virasoro algebra, $J$ is the conformal charge defined by $J:=\mathbf{rt}%
\det (\mathbf{R})$, and $\mathbf{R}$ is the $m\times m$ symmetric positive
definite compactification matrix of the system of $m$ free boson fields (see
section 4 for details).

The h.w.\ $\Lambda $ of $su(m)$ in $\left( \ref{ch-w-m}\right) $ is defined
by $\left( \ref{def-Lambda}\right) $ in terms of the h.w. $\mathbf{r\in }%
\mathbb{P}^{(m)}$ and the identity $\left( \ref{ch-w-m}\right) $ follows as
a consequence of the following relation between the conformal dimensions $h_{%
\mathbf{r}}$ and $h_{\Lambda }$: 
\begin{equation}
h_{\mathbf{r}}=\frac{1}{2m}\left( \sum_{i=1}^{m}r_{i}\right) ^{2}+h_{\Lambda
}\text{,}
\end{equation}
\vspace{0.1in}where: 
\begin{equation}
\hspace{0.05in}h_{\mathbf{r}}:=\frac{1}{2}\sum_{i=1}^{m}r_{i}^{2}\hspace{%
0.05in}\text{ \ and \ \ }h_{\Lambda }:=\frac{\left| \Lambda \right| ^{2}}{2}=%
\frac{1}{2}\sum_{i=1}^{m}\left( r_{i}-\frac{1}{m}\sum_{j=1}^{m}r_{j}\right)
^{2}\text{.}
\end{equation}

\setcounter{equation}{0}

\section{\label{main results}The main identities}

In this section, we derive the main results of the paper that allow us to
show that our theory TM gives a new $\Gamma _{\theta }$-RCFT extension of
the fully degenerate $W_{1+\infty }^{(m)}$.\smallskip

\begin{prop}
\label{prop 1}\textit{Let }$\chi _{\widehat{\Lambda }_{l}}^{\widehat{su(m)}%
_{1}}(\xi |\tau )$\textit{\ be the character of the affine level 1 }$%
\widehat{su(m)}_{1}$ \textit{corresponding to the fundamental representation 
}$\widehat{\Lambda }_{l}$\textit{, then it has the following series
expansion in terms of the characters of }$\mathcal{W}_{m}$: 
\begin{equation}
\chi _{\widehat{\Lambda }_{l}}^{\widehat{su(m)}_{1}}(\xi =\frac{\rho }{m}%
|\tau )=\delta _{l,0}\sum_{\Lambda \in D_{m}}\epsilon (w_{\Lambda })\chi
_{\Lambda }^{\mathcal{W}_{m}}(\tau )\text{,}  \label{F-2}
\end{equation}
\textit{where }$D_{m}:=\{\Lambda \in P_{+}:\exists !w_{\Lambda }\in
W\rightarrow w_{\Lambda }(\Lambda +\rho )-\rho \in mQ\}$\textit{, }$\epsilon
(w_{\Lambda })\ $\textit{is the signature of the Weyl reflection} $%
w_{\Lambda }$\textit{,}$\ P_{+}$\textit{\ is the set of the dominant weights
and }$Q$\textit{\ is the root lattice of the finite algebra }$su(m)$\textit{.%
}\vspace{0.07in}

\textit{Let }$\eta (\tau )$\textit{\ be the Dedekind function, then it holds:%
} 
\begin{equation}
\chi _{\widehat{\Lambda }_{l}}^{\widehat{su(m)}_{1}}(\xi =\frac{\rho }{m}%
|\tau )=\delta _{l,0}\frac{\eta (\tau )}{\eta (m\tau )}\text{.}  \label{F-1}
\end{equation}
\vspace{0.15in}
\end{prop}

To prove this Proposition we make use of some important results of B.
Kostant already given in \cite{Kostant}. Here, we just recall those results
which turn out to be useful for us and we restate them according to our
notations.\vspace{0.1in}

\begin{Kprop}
\textit{Let g be a simple Lie algebra. Then, either}

\textit{(1) }$\forall w\in W\rightarrow w(\Lambda +\rho )-\rho \notin hQ$%
\textit{, or}

\textit{(2) }$\exists !w_{\Lambda }\in W\rightarrow w_{\Lambda }(\Lambda
+\rho )-\rho \in hQ$\textit{,\ where }$h$\textit{\ is the Coxeter number of}
g\textit{.\vspace{0.1in}}
\end{Kprop}

This Proposition, in particular, makes clear the definition given above of
the set $D_{m}$, if one recalls that $m$ is the Coxeter number of $su(m)$.
Kostant defines the so called principal element of type $\rho $ and here we
give a definition of it in a way to be independent of the normalization of
the Killing form.\vspace{0.1in}

\begin{Def}
\label{def}\textit{Let }g\textit{\ be a simply laced Lie algebra and }$x_{p}$
\textit{be defined as an element of }g\textit{\ such that }$\left(
x_{p},\alpha _{i}\right) :=1/h$\textit{, }$i\in \{1,..,h-1\}$\textit{, where 
}$\alpha _{i}$\textit{\ are the simple roots, }$(\hspace{0.04in},)$\textit{\
is the Killing form and }$h$\textit{\ is the Coxeter number of }g\textit{.}

\textit{Then, every element of }g\textit{\ conjugate, with respect to the
Weyl group, to the element }$x_{p}$ \textit{is called principal of type }$%
\rho $\textit{.\vspace{0.1in}}
\end{Def}

The normalization of the Killing form that we chose is $\left( \alpha
_{i},\alpha _{i}\right) =2$, $i\in \{1,..,h-1\}$, in this case $\left( \rho
,\alpha _{i}\right) =1$ and thus $x_{p}=\rho /h$.\vspace{0.1in}

\begin{Kprop}
\textit{For any }$\Lambda \in P_{+}$\textit{\ and }$\xi $\textit{\ principal
of type }$\rho $\textit{\ it results }$\chi _{\Lambda }^{\text{g}}(\xi )\in
\{-1,0,1\}$\textit{, where }$\chi _{\Lambda }^{\text{g}}$\textit{\ is the
character of the representation }$\Lambda $\textit{. In particular, it
holds: } 
\begin{equation}
\chi _{\Lambda }^{\text{g}}(\xi )=\left\{ 
\begin{array}{c}
0\hspace{0.05in}for\hspace{0.05in}\Lambda \notin D_{h} \\ 
\epsilon (w_{\Lambda })\hspace{0.05in}for\hspace{0.05in}\Lambda \in D_{h}
\end{array}
\right. \text{,}
\end{equation}
\textit{where }$D_{h}:=\{\Lambda \in P_{+}:\exists !w_{\Lambda }\in
W\rightarrow w_{\Lambda }(\Lambda +\rho )-\rho \in hQ\}$\textit{.}\vspace{%
0.1in}
\end{Kprop}

In the case of $su(m)$, then $x_{p}=\rho /m$ is a principal element of type $%
\rho $ and the following identities hold: 
\begin{equation}
\chi _{\Lambda }^{su(m)}(\xi =\frac{\rho }{m})=\left\{ 
\begin{array}{c}
0\hspace{0.05in}\forall \hspace{0.05in}\Lambda \notin D_{m} \\ 
\epsilon (w_{\Lambda })\hspace{0.05in}\forall \hspace{0.05in}\Lambda \in
D_{m}
\end{array}
\right. .  \label{su(m)-ro/m}
\end{equation}
We give now a first characterization of the weights of $D_{m}$ in terms of
the finite part of the affine weights of the fundamental representations.%
\vspace{0.1in}

The finite part\footnote{%
The $m$-ality of an affine weight $\widehat{\Lambda }=[\lambda _{0},\lambda
_{1},..,\lambda _{m-1}]$ is defined as the $m$-ality of its finite part $%
\Lambda =[\lambda _{1},..,\lambda _{m-1}]$. So, Lemma \ref{lem 1} implies
that the affine weight system $\widehat{\Omega }_{l}$ has the same $m$-ality
of the finite part $\Omega _{l}$, i.e. $l/m$.} $\Omega _{l}$ of the affine
weight systems $\widehat{\Omega }_{l}$, $l\in \{0,..,m-1\}$,\ can be
characterized by the so called $m$-ality, or congruence class, of the
weights.\vspace{0.1in}

\begin{lemma}
\label{lem 1}\textit{The $m$-ality of the weight }$\Lambda =[\lambda
_{1},..,\lambda _{m-1}]$\textit{, where }$\lambda _{i}$\textit{\ are the
Dynkin labels, is defined as }$k(\Lambda )=\frac{1}{m}\sum_{i=1}^{m-1}i%
\lambda _{i}$\textit{\ mod}$1$\textit{. Then, all the weights in }$\Omega
_{l}$\textit{\ have the same $m$-ality }$l/m$\textit{, }$l\in \{0,..,m-1\}$%
\textit{.}
\end{lemma}

\textit{Proof ---}\ \ We have to prove that $k(\Lambda )=l/m$ $\forall
\Lambda \in \Omega _{l}.$ This is of course true for the finite fundamental
weights $\Lambda _{l}$, that is $k(\Lambda _{l})=l/m.$ Now, the generic
weight $\Lambda \in \Omega _{l}$ has the form $\Lambda =\Lambda
_{l}+\sum_{i=1}^{m-1}n_{i}\alpha _{i}$, and by using $\alpha _{i}=2\Lambda
_{i}-(1-\delta _{i,1})\Lambda _{i-1}-(1-\delta _{i,m-1})\Lambda _{i+1}$ we
get $k(\alpha _{i})=\frac{\left( i+1\right) }{m}\delta _{i,m-1}=0$ mod$1$.
Thus, the equalities $k(\Lambda )=k(\Lambda _{l})=l/m$ hold.

\hfill $\Box$

\vspace{0.15in}

\begin{lemma}
\label{lem 2}All the weights of $D_{m}$ have zero $m$-ality, that is: $%
D_{m}\subset P_{+}\cap \Omega _{0}$.
\end{lemma}

\textit{Proof ---}\ \ If $\Lambda \in D_{m}$ then $w_{\Lambda }(\Lambda
+\rho )-\rho \in mQ$, by definition for each element of the Weyl group, $%
\exists !n_{i}\in \mathbb{Z}:$ $w_{\Lambda }(\Lambda +\rho )=\Lambda +\rho
+\sum_{i=1}^{m-1}n_{i}\alpha _{i}$. The condition $\Lambda \in D_{m}$ so
implies that $\exists !q_{i}\in m\mathbb{Z}:$ $\Lambda
+\sum_{i=1}^{m-1}n_{i}\alpha _{i}=\sum_{i=1}^{m-1}q_{i}\alpha _{i}$, that is 
$\exists !p_{i}\in \mathbb{Z}:\Lambda =\sum_{i=1}^{m-1}p_{i}\alpha _{i}$.
Thus, $\Lambda $ has zero $m$-ality and $\Lambda \in P_{+}\cap \Omega _{0}$.

\hfill $\Box$

\vspace{0.1in}

Thus, we can give the following \textit{proof of the identity }$\left( \ref
{F-2}\right) $\textit{\ ---}\ \ The identity $\left( \ref{F-2}\right) $ is
an immediate consequence of Lemma \ref{lem 2}, $\left( \ref{su(m)-ro/m}%
\right) $ and $\left( \ref{coset-1}\right) $ evaluated at $\xi =\rho /m$.

\hfill $\Box$

\vspace{0.15in}

We are now ready to prove the identity $\left( \ref{F-1}\right) $. Let us
start by giving the following Lemma.\vspace{0.06in}

\begin{lemma}
\label{lem 3}\textit{The ratio }$\eta (\tau )/\eta (m\tau )$\textit{\ has
the equivalent expression in terms of }$\Theta $\textit{-functions:} 
\begin{equation}
\frac{\eta (\tau )}{\eta (m\tau )}=\frac{1}{\eta (\tau )^{m-1}}\frac{%
\prod_{j=1}^{m}\Theta _{3}\left( w-\frac{j}{m}|\tau \right) }{\Theta
_{3}\left( mw-\frac{m+1}{2}|m\tau \right) }\text{.}  \label{ratio of neta}
\end{equation}
\end{lemma}

\textit{Proof ---}\ \ The identity $\left( \ref{ratio of neta}\right) $ is
equivalent to the following one: 
\begin{equation}
\frac{\Theta _{3}\left( mw-\frac{m+1}{2}|m\tau \right) }{\eta (m\tau )}%
=\prod_{j=1}^{m}\frac{\Theta _{3}\left( w-\frac{j}{m}|\tau \right) }{\eta
(\tau )}\text{.}  \label{ratio of
neta-2}
\end{equation}
Using the definition of $\eta (\tau )$ and the expansion of $\Theta
_{3}\left( w|\tau \right) $ in terms of infinite products: 
\begin{equation}
\eta (\tau )=q^{1/24}\prod_{n=1}^{\infty }(1-q^{n}),\hspace{0.1in}\Theta
_{3}\left( w|\tau \right) =q^{-1/24}\eta (\tau )\prod_{n=1}^{\infty
}(1-yq^{n-1/2})(1-y^{-1}q^{n-1/2})\text{,}
\end{equation}
where $q:=e^{2\pi i\tau }$ and $y:=e^{2\pi iw}$, we obtain: 
\begin{align}
& \frac{\Theta _{3}\left( mw-\frac{m+1}{2}|m\tau \right) }{\eta (m\tau )}%
=q^{-m/24}\prod_{n=1}^{\infty }(1-(-1)^{\frac{m+1}{2}%
}y^{m}q^{m(n-1/2)})(1-(-1)^{\frac{m+1}{2}}y^{-m}q^{m(n-1/2)}) \\
& \mbox{and}  \notag \\
& \prod_{j=1}^{m}\frac{\Theta _{3}\left( w-\frac{j}{m}|\tau \right) }{\eta
(\tau )}=q^{-m/24}\prod_{j=1}^{m}\prod_{n=1}^{\infty }(1-e^{-\frac{2\pi i}{m}%
j}yq^{n-1/2})(1-e^{-\frac{2\pi i}{m}j}y^{-1}q^{n-1/2})\text{.}
\end{align}
The identity in $\left( \ref{ratio of neta-2}\right) $ is then an immediate
consequence of the following one: 
\begin{equation}
\prod_{j=1}^{m}\left( 1+ae^{-\frac{2\pi i}{m}j}\right) =\left( 1+(-1)^{\frac{%
m+1}{2}}a^{m}\right) ,
\end{equation}
due to the properties of the roots of the unity.

\hfill $\Box $

\vspace{0.15in}

Furthermore, by using the series expansion of $\Theta _{3}\left( w|\tau
\right) $, it holds: 
\begin{equation}
\prod_{j=1}^{m}\Theta _{3}\left( w-\frac{j}{m}|\tau \right) =\sum_{\left(
n_{1},..,n_{m}\right) \in \mathbb{Z}^{m}}q^{\frac{1}{2}%
\sum_{i=1}^{m}n_{i}^{2}}e^{2i\pi \sum_{j=1}^{m}(w-\frac{j}{m})n_{j}}\text{.}
\label{teta3^m}
\end{equation}
If we define $n_{X}:=\left( \sum_{i=1}^{m}n_{i}\right) /m$ and $%
u_{j}:=n_{j}-n_{X}$, then: 
\begin{equation}
n_{X}:=n_{T}+l/m,  \label{nx}
\end{equation}
with $l\in \{0,..,m-1\}$, $n_{T}\in \mathbb{Z}$ and $\sum_{j=1}^{m}u_{j}=0$.
The last condition makes possible to interpret $\Lambda
=\sum_{i=1}^{m}u_{i}\epsilon _{i}$ as a weight of $su(m)$, where $\epsilon
_{1},..,\epsilon _{m}$ is an orthonormal basis of the Euclidean space $%
\mathbb{R}^{m}$ \cite{wyb,FH,Di Francesco}. In terms of Dynkin labels $%
\Lambda $ can be rewritten as $\Lambda =\sum_{j=1}^{m-1}\lambda _{j}\Lambda
_{j}$, where $\lambda _{j}:=u_{j}-u_{j+1}=n_{j}-n_{j+1}$, $j\in \{1,..,m-1\}$%
, are integer numbers. Furthermore, $\left( \sum_{i=1}^{m}i\lambda
_{i}\right) /m=$ $n_{T}-n_{m}+l/m$ and thus the $m$-ality of the weight $%
\Lambda $ coincides with $l/m$ in $\left( \ref{nx}\right) $ and $\Lambda \in
\Omega _{l}$. The sums in $\left( \ref{teta3^m}\right) $ can then be
rewritten as: 
\begin{equation}
\frac{1}{2}\sum_{i=1}^{m}n_{i}^{2}=\frac{m}{2}n_{X}^{2}+\frac{1}{2}%
\sum_{i=1}^{m}u_{i}^{2}=\frac{m}{2}n_{X}^{2}+h_{\Lambda }\text{,}
\end{equation}
where: 
\begin{equation}
h_{\Lambda }:=\frac{1}{2}\left| \Lambda \right| ^{2}=\frac{1}{2}%
\sum_{i=1}^{m}u_{i}^{2}  \label{h-lambda}
\end{equation}
and 
\begin{equation}
\sum_{j=1}^{m}(w-\frac{j}{m})n_{j}=n_{X}(mw-\frac{m+1}{2})-\frac{1}{m}%
\sum_{j=1}^{m}ju_{j}\text{.}
\end{equation}
By recalling that, in terms of the fundamental weights of $su(m)$, $\rho $
has the expansion $\rho =\sum_{j=1}^{m-1}\Lambda _{j}$ one gets in the $%
\epsilon _{1},..,\epsilon _{m}$ basis: 
\begin{equation}
\rho =\sum_{i=1}^{m}\left( \frac{m+1}{2}-i\right) \epsilon _{i},
\end{equation}
and so: 
\begin{equation}
\frac{1}{m}\left( \Lambda ,\rho \right) =\frac{1}{m}\sum_{j=1}^{m}\left( 
\frac{m+1}{2}-j\right) u_{j}=-\frac{1}{m}\sum_{j=1}^{m}ju_{j}\text{.}
\end{equation}
Using the above identities, $\left( \ref{teta3^m}\right) $ can be rewritten
as: 
\begin{equation}
\prod_{v=1}^{m}\Theta _{3}\left( w-\frac{v}{m}|\tau \right)
=\sum_{l=0}^{m-1}\left\{ \left[ \sum_{n_{T}\in \mathbb{Z}}q^{\frac{m}{2}%
(n_{T}+\frac{l}{m})^{2}}e^{2i\pi (mw-\frac{m+1}{2})(n_{T}+\frac{l}{m})}%
\right] \left[ \sum_{\Lambda \in \Omega _{l}}q^{h_{\Lambda }}e^{2i\pi \left(
\Lambda ,\frac{\rho }{m}\right) }\right] \right\} \text{.}
\end{equation}
Furthermore, by expressing this last factor in terms of $\Theta $-functions
with characteristics: 
\begin{equation}
\Theta \left[ 
\begin{array}{c}
a \\ 
b
\end{array}
\right] \left( w|\tau \right) :=\sum_{u\in \mathbb{Z}}e^{\pi i\tau \left(
u+a\right) ^{2}+2\pi i(w+b)\left( u+a\right) }
\end{equation}
and substituting the result in $\left( \ref{ratio of neta}\right) $, it
holds: 
\begin{equation}
\frac{\eta (\tau )}{\eta (m\tau )}=\sum_{l=0}^{m-1}G_{l}^{(m)}(\tau )\left( 
\frac{1}{\eta (\tau )^{m-1}}\sum_{\Lambda \in \Omega _{l}}q^{h_{\Lambda
}}e^{2i\pi \left( \Lambda ,\frac{\rho }{m}\right) }\right) \text{,}
\end{equation}
where: 
\begin{equation}
G_{l}^{(m)}(\tau ):=\left. \Theta \left[ 
\begin{array}{c}
\frac{l}{m} \\ 
\frac{-(m+1)}{2}
\end{array}
\right] \left( mw|m\tau \right) \right/ \Theta \left[ 
\begin{array}{c}
0 \\ 
\frac{-(m+1)}{2}
\end{array}
\right] \left( mw|m\tau \right) \text{.}
\end{equation}
V. Kac and D. Peterson \cite{KP2} have shown that for any level 1 simply
laced affine Lie algebra $X_{a}^{(1)}$, all the non-zero string functions
coincide with $1/\eta (\tau )^{a}$ and thus the characters of the
fundamental representations of $\widehat{su(m)}_{1}$ read as: 
\begin{equation}
\chi _{\widehat{\Lambda }_{l}}^{\widehat{su(m)}_{1}}(\xi |\tau )=\frac{1}{%
\eta (\tau )^{m-1}}\sum_{\Lambda \in \Omega _{l}}q^{h_{\Lambda }}e^{2i\pi
\left( \Lambda ,\xi \right) },  \label{f-K-P}
\end{equation}
and we finally get: 
\begin{equation}
\frac{\eta (\tau )}{\eta (m\tau )}=\sum_{l=0}^{m-1}G_{l}^{(m)}(\tau )\chi _{%
\widehat{\Lambda }_{l}}^{\widehat{su(m)}_{1}}(\xi =\frac{\rho }{m}|\tau )%
\text{.}
\end{equation}
But from $\left( \ref{F-2}\right) $, we know that $\chi _{\widehat{\Lambda }%
_{l}}^{\widehat{su(m)}_{1}}(\xi =\frac{\rho }{m}|\tau )$ is non-zero only
for $l=0$ and thus $\left( \ref{F-1}\right) $ immediately follows, so ending
the proof of Proposition \ref{prop 1}.

\hfill $\Box$

The results given in Proposition \ref{prop 1} are essential in order to
define an extension of the chiral algebra $\mathcal{W}_{m}$.

We have shown that $D_{m}$ is contained in $\Omega _{0}$, however $\Omega
_{0}$ has infinite weights and so it is important to look for a simpler
definition of $D_{m}$. In the following, we will give a remarkable
simplification characterizing $D_{m}$ in terms of a finite subset of $\Omega
_{0}$.

Let $P_{m,+}$ be the subset of $P_{+}$ whose weights have Dynkin labels in $%
\{0,..,m-1\}$, so $P_{m,+}$ is a finite subset of $P_{+}$ and $P_{+}=$\ $%
P_{m,+}+mP_{+}$. That is, any weight $\Lambda \mathbf{\in }P_{+}$ has the
form $\Lambda ^{\prime }+m\Lambda ^{\prime \prime }$ where $\Lambda ^{\prime
}\in P_{m,+}$ and $\Lambda ^{\prime \prime }\in P_{+}$, $\Lambda ^{\prime }$
being the module $m$ part of $\Lambda $.\vspace{0.1in}

\begin{prop}
\label{prop 2}\textit{The identities of Proposition \ref{prop 1} can be
written in the following more explicit forms:}

\textit{For }$m$\textit{\ odd: } 
\begin{equation}
\chi _{\widehat{\Lambda }_{l}}^{\widehat{su(m)}_{1}}(\xi =\frac{\rho }{m}%
|\tau )=\delta _{l,0}\frac{\eta (\tau )}{\eta (m\tau )}=\sum_{\Lambda
^{\prime \prime }\in P_{+}}\hspace{0.05in}\hspace{0.05in}\sum_{\Lambda
^{\prime }\in P_{m,+}\cap D_{m}}\epsilon (w_{\Lambda ^{\prime }})\chi
_{\Lambda ^{\prime }+m\Lambda ^{\prime \prime }}^{\mathcal{W}_{m}}(\tau )%
\text{.}  \label{F-simply-odd}
\end{equation}

\textit{For }$m=2n$\textit{\ even:} 
\begin{equation}
\chi _{\widehat{\Lambda }_{l}}^{\widehat{su(m)}_{1}}(\xi =\frac{\rho }{m}%
|\tau )=\delta _{l,0}\frac{\eta (\tau )}{\eta (m\tau )}=\sum_{\Lambda
^{\prime \prime }\in P_{+}}\hspace{0.05in}\hspace{0.05in}\sum_{\Lambda
^{\prime }\in P_{m,+}\cap D_{m}}(-1)^{\sum_{i=0}^{n-1}\lambda
_{2i+1}^{\prime \prime }}\epsilon (w_{\Lambda ^{\prime }})\chi _{\mathbf{\ }%
\Lambda ^{\prime }+m\Lambda ^{\prime \prime }}^{\mathcal{W}_{m}}(\tau )\text{%
,}  \label{F-simply-even}
\end{equation}
\textit{where }$\lambda _{i}^{\prime \prime }$\textit{\ are the Dynkin
labels of }$\Lambda ^{\prime \prime }$\textit{.\vspace{0.1in}}
\end{prop}

To prove Proposition \ref{prop 2} we start proving the Lemma. \vspace{0.1in}

\begin{lemma}
\label{lem 4}\textit{The following equivalent characterization of }$%
D_{m}=\{\Lambda \in P_{+}:\Lambda =\Lambda ^{\prime }+m\Lambda ^{\prime
\prime }\hspace{0.1in}$\textit{with }$\Lambda ^{\prime }\in P_{m,+}\cap D_{m}%
\hspace{0.1in}$\textit{and}$\hspace{0.1in}\Lambda ^{\prime \prime }\in
P_{+}\}$\textit{\ holds. Furthermore, for every weight }$\Lambda \in P_{+}$%
\textit{\ of the form }$\Lambda ^{\prime }+m\Lambda ^{\prime \prime }$%
\textit{, where }$\Lambda ^{\prime }\in P_{m,+}$\textit{\ and }$\Lambda
^{\prime \prime }\in P_{+}$\textit{, it results: } 
\begin{equation}
\chi _{\Lambda }^{su(m)}(\xi =\frac{\rho }{m})=\chi _{\mathbf{\ }\Lambda
^{\prime }}^{su(m)}(\xi =\frac{\rho }{m})\hspace{0.2in}for\hspace{0.06in}m%
\hspace{0.06in}odd  \label{Kostant-simply-odd}
\end{equation}
and 
\begin{equation}
\chi _{\Lambda }^{su(m)}(\xi =\frac{\rho }{m})=(-1)^{\sum_{i=0}^{n-1}\lambda
_{2i+1}^{\prime \prime }}\text{ }\chi _{\Lambda ^{\prime }}^{su(m)}(\xi =%
\frac{\rho }{m})\hspace{0.2in}for\hspace{0.06in}m=2n\hspace{0.06in}even.
\label{Kostant-simply-even}
\end{equation}
\end{lemma}

\textit{Proof ---\ \ }Kostant's Proposition 2 implies that $D_{m}$ is the
subset of the dominant weights $P_{+}$ that give a non-zero value of $\chi
_{\Lambda }^{su(m)}(\xi =\frac{\rho }{m})$. Thus, the above characterization
of $D_{m}$ follows by the proof of the identities $\left( \ref
{Kostant-simply-odd}\right) $ and $\left( \ref{Kostant-simply-even}\right) $%
. Indeed, they imply that $\chi _{\Lambda }^{su(m)}(\xi =\frac{\rho }{m}%
)\neq 0$ if and only if $\chi _{\Lambda ^{\prime }}^{su(m)}(\xi =\frac{\rho 
}{m})\neq 0$, that is $\Lambda \mathbf{(=}\Lambda ^{\prime }+m\Lambda
^{\prime \prime })\in D_{m}$ if and only if $\Lambda ^{\prime }\in
P_{m,+}\cap D_{m}$ and $\Lambda ^{\prime \prime }\in P_{+}$. Whenever the
character of $su(m)$ is evaluated at $\xi \varpropto \rho $, it is possible
to use the following expression: 
\begin{equation}
\chi _{\Lambda }^{su(m)}(\xi =\frac{\rho }{m})=\prod\limits_{\alpha >0}\frac{%
\sin \frac{\pi }{m}(\alpha {,}(\Lambda +\rho ))}{\sin \frac{\pi }{m}(\alpha {%
,}\rho )}=\prod\limits_{i=1}^{m-1}\hspace{0.05in}\prod\limits_{h=0}^{m-1-i}%
\frac{\sin \frac{\pi }{m}(\lambda _{i}+...+\lambda _{i+h}+h+1)}{\sin \frac{%
\pi }{m}(h+1)}\text{.}
\end{equation}
Writing $\Lambda \mathbf{=}\Lambda ^{\prime }+m\Lambda ^{\prime \prime }$
and using the expansion of sin$(a+b)$, it follows: 
\begin{equation}
\chi _{\Lambda }^{su(m)}(\xi =\frac{\rho }{m})=(-1)^{a(m,\mathbf{\ }\Lambda
^{\prime \prime })}\chi _{\Lambda ^{\prime }}^{su(m)}(\xi =\frac{\rho }{m})%
\text{,}  \label{Kostant-simply}
\end{equation}
where $a(m,\Lambda ^{\prime \prime
})=\sum_{i=1}^{m-1}\sum_{h=0}^{m-1-i}(\lambda _{i}^{\prime \prime
}+...+\lambda _{i+h}^{\prime \prime })$, in terms of the Dynkin labels $%
[\lambda _{1}^{\prime \prime },..,\lambda _{m-1}^{\prime \prime }]$ of $%
\Lambda ^{\prime \prime }$. Such exponent can be rewritten as $a(m,\Lambda
^{\prime \prime })=\sum_{i=1}^{m-1}n(m,i)\lambda _{i}^{\prime \prime }$,
where $n(m,i):=i(m-i)$.

Thus, for $m$ odd all the $n(m,i)$ are even integers, while for $m$ even $%
n(m,i)$ are even for $i$ even and odd for $i$ odd. So, $\left( \ref
{Kostant-simply}\right) $ implies $\left( \ref{Kostant-simply-odd}\right) $
and $\left( \ref{Kostant-simply-even}\right) $, ending the proof of Lemma 
\ref{lem 4}.

\hfill $\Box$

\vspace{0.15in}

It is worth pointing out that $P_{m,+}\cap D_{m}$ is a finite subset of the
dominant weights with zero $m$-ality, which implies the announced
simplification in the characterization of $D_{m}$.\vspace{0.1in}

\textit{Proof of Proposition \ref{prop 2} --- \ }The proof is now an
immediate consequence of Proposition \ref{prop 1} and Lemma \ref{lem 4}.
Indeed, by substituting identities $\left( \ref{Kostant-simply-odd}\right) $
and $\left( \ref{Kostant-simply-even}\right) $ in $\left( \ref{F-2}\right) $%
, the equations $\left( \ref{F-simply-odd}\right) $ and $\left( \ref
{F-simply-even}\right) $ of Proposition \ref{prop 2} follow.

\hfill $\Box$

\vspace{0.15in}

Finally, we use the identity $\left( \ref{F-1}\right) $ to derive the last
result of this section.\vspace{0.1in}

\begin{cor}
\label{cor}\textit{The following identity holds:} 
\begin{equation}
\frac{\eta (\tau )}{\eta (\tau /m)}=F_{twist}^{(m)}(\tau )\sum_{l=0}^{m-1}%
\mathbf{S}_{0,l}^{\widehat{su(m)}_{1}}\chi _{\widehat{\Lambda }_{l}}^{%
\widehat{su(m)}_{1}}(\xi =\frac{\rho }{m}\tau |\tau )\text{,}
\label{twistato neta ratio}
\end{equation}
\textit{where }$\mathbf{S}_{a,b}^{\widehat{su(m)}_{1}}$\textit{, }$a,b\in
\{0,..,m-1\},$\textit{\ are the elements of the unitary matrix }$\mathbf{S}^{%
\widehat{su(m)}_{1}}$\textit{\ that define the action of the modular
transformation }$S:$\textit{\ }$\tau \rightarrow -1/\tau $\textit{\ on the
characters of }$\widehat{su(m)}_{1}$\textit{.}

\textit{The explicit expression is }\cite{Di Francesco}\textit{: } 
\begin{equation}
\mathbf{S}_{a,b}^{\widehat{su(m)}_{1}}=i^{\left| \Delta _{+}\right| }\left(
\det A^{su(m)}\right) ^{-1/2}(1+m)^{-(m^{2}-1)}\sum_{w\epsilon W}\epsilon
(w)e^{-2\pi i\left( w(\Lambda _{a}+\rho ){,}\Lambda _{b}+\rho \right) /(m+1)}%
\text{,}
\end{equation}
\textit{where }$\left| \Delta _{+}\right| $\textit{\ is the number of
positive roots, }$A^{su(m)}$\textit{\ is the Cartan matrix, }$W$\textit{\ is
the Weyl group, }$\Lambda _{a}$\textit{\ and }$\Lambda _{b}$\textit{\ are
the fundamental weights of }$su(m)$\textit{\ and } 
\begin{equation}
F_{twist}^{(m)}(\tau )=\frac{1}{\sqrt{m}}e^{2\pi i\left( \frac{m^{2}-1}{24m}%
\right) \tau }.  \label{f-twist}
\end{equation}
\vspace{0.1in}
\end{cor}

\textit{Proof ---\ \ }The Corollary is a direct consequence of $\left( \ref
{F-1}\right) $ and of the well known action of the modular transformation $%
S: $ $\tau \rightarrow -1/\tau $ on the characters of $\widehat{su(m)}_{1}$
and on $\eta (\tau )$. By definition of $S$ and by using $\left( \ref{F-1}%
\right) $, it results: $S(\eta (\tau )/\eta (m\tau )):=\eta (-1/\tau )/\eta
(-m/\tau )$ and $S(\eta (\tau )/\eta (m\tau ))=\chi _{\widehat{\Lambda }%
_{0}}^{\widehat{su(m)}_{1}}(\xi =\rho /m|-1/\tau )$, respectively. Expanding
the right hand side of these last two equalities, we obtain our result $%
\left( \ref{twistato neta ratio}\right) $ with the following expression for $%
F_{twist}^{(m)}(\tau )$: 
\begin{equation}
F_{twist}^{(m)}(\tau )=\frac{1}{\sqrt{m}}e^{2\pi i(\frac{1}{2}\tau \left|
\rho \right| ^{2})/m^{2}}\text{.}
\end{equation}
Finally, by using the Freudental-de Vries strange formula $\left| \rho
\right| ^{2}$=$\left( m/12\right) \dim su(m)$=$m(m^{2}-1)/12$, $%
F_{twist}^{(m)}(\tau )$ takes the expression $\left( \ref{f-twist}\right) $.

\hfill $\Box$

\vspace{0.15in}

\setcounter{equation}{0}

\section{The $\Gamma _{\protect\theta }$-RCFT $\widehat{u(1)}_{\mathbf{K}%
_{m,p}}$}

In order to make more clear the derivation of our theory TM, we introduce
here the $\Gamma _{\theta }$-RCFT $\widehat{u(1)}_{\mathbf{K}_{m,p}}$.

As it is well known, the $m$-component free boson $\widehat{u(1)}^{\otimes
m} $, with chiral algebra $\mathfrak{A}(\widehat{u(1)})^{\otimes m}$ (the
tensor product of $m$ Heisenberg algebras), is not a rational CFT. Let $%
\mathbf{K}_{m,p}$ be the $m\times m$ symmetric matrix with integer entries: 
\begin{equation}
\mathbf{K}_{m,p}=\mathbf{1}_{m{\times }m}+2p\mathbf{C}_{m{\times }m},
\end{equation}
where $p\in \mathbb{Z}$ and $\mathbf{C}_{m{\times }m}$ is the $m\times m$
matrix, all elements of which are equal to 1. An RCFT can be defined now
imposing on $\widehat{u(1)}^{\otimes m}$ the following compactification
condition fixed by $\mathbf{K}_{m,p}$ \cite{WZ,MS}: 
\begin{equation}
\mathbf{\varphi }(ze^{2\pi i},\bar{z}e^{-2\pi i})=\mathbf{\varphi }(z,\bar{z}%
)+2\pi \mathbf{R}_{m,p}\mathbf{h}\text{,}  \label{compattificazione}
\end{equation}
where $\mathbf{h}^{T}:=(h_{1},...,h_{m})\in \mathbb{Z}^{m}$ is the winding
vector, $\mathbf{\varphi }(z,\bar{z})\ $is defined\ by $\mathbf{\varphi }(z,%
\bar{z})^{T}:=(\varphi ^{(1)}(z,\bar{z}),$\newline
$...,\varphi ^{(m)}(z,\bar{z}))$, with $\varphi ^{(i)}(z,\bar{z})$ free
boson fields, and $\mathbf{R}_{m,p}$ is the $m{\times }m$ matrix defined by%
\footnote{$\mathbf{R}_{m,p}$ is the positive root of $\mathbf{K}_{m,p}$,
which is well defined because $\mathbf{K}_{m,p}$ is a symmetric and positive
definite $m{\times }m$ matrix.}: 
\begin{equation}
\mathbf{R}_{m,p}^{T}\mathbf{R}_{m,p}=\mathbf{K}_{m,p}\text{,}
\end{equation}
that explicitly reads: 
\begin{equation}
\mathbf{R}_{m,p}=\mathbf{1}_{m{\times }m}+\frac{1}{m}\left( \sqrt{2pm+1}%
-1\right) \mathbf{C}_{m{\ {\times }}m}\text{\hspace{0.01in}.}
\end{equation}
The compactification condition $(\ref{compattificazione})$ for diagonal $%
\mathbf{h}=h\mathbf{t}$ defines the following ones: 
\begin{equation}
\varphi ^{(i)}(ze^{2\pi i},\bar{z}e^{-2\pi i})=\varphi ^{(i)}(z,\bar{z}%
)+2\pi rh,  \label{comp-free-bos-i}
\end{equation}
for the free boson fields $\varphi ^{(i)}(z,\bar{z})$ with$\,h\in \mathbb{Z}%
, $ where the square of the compactification radius $r$ is an odd number, $%
r^{2}=2pm+1$.

The compactification condition has the effect to influence the zero-modes $%
\mathbf{a}_{0}:=(a_{0}^{(1)},...,a_{0}^{(m)})$ of the free boson fields
only. In particular, to obtain well defined vertex operators, under the
compactification condition, the possible eigenvalues of $\mathbf{a}_{0}$ are
restricted to the following values $\mathbf{\alpha }_{\mathbf{p}}=\mathbf{p}%
^{T}\mathbf{R}_{m,p}^{-1}$ with $\mathbf{p}\in \mathbb{Z}^{m}$ and 
\begin{equation}
\mathbf{R}_{m,p}^{-1}=\mathbf{1}_{m{\times }m}+\frac{1}{m}\left( \frac{1}{%
\sqrt{2pm+1}}-1\right) \mathbf{C}_{m{\ {\times }}m}\text{.}
\end{equation}
So, the h.w. vectors of the compactified $\widehat{u(1)}^{\otimes m}$ are $%
\left| \mathbf{\alpha }_{\mathbf{p}}^{{}}\right\rangle $:=$%
\bigotimes_{i=1}^{m}\left| \mathbf{\alpha }_{\mathbf{p}}^{\left( i\right)
}\right\rangle $, where $\mathbf{\alpha }_{\mathbf{p}}$=$(\mathbf{\alpha }_{%
\mathbf{p}}^{\left( 1\right) },...,\mathbf{\alpha }_{\mathbf{p}}^{\left(
m\right) })$ and $\left| \mathbf{\alpha }_{\mathbf{p}}^{\left( i\right)
}\right\rangle $ are the h.w. vectors of the $i^{th}$ Heisenberg algebra $%
\{a_{k}^{(i)}\}_{k\in \mathbb{Z}}$, corresponding to the eigenvalue $\left( 
\mathbf{\alpha }_{\mathbf{p}}\right) ^{\left( i\right) }$ of $a_{0}^{(i)}$,
i.e. 
\begin{equation}
a_{0}^{(i)}\left| \mathbf{\alpha }_{\mathbf{p}}^{\left( i\right)
}\right\rangle =\mathbf{\alpha }_{\mathbf{p}}^{\left( i\right) }\left| 
\mathbf{\alpha }_{\mathbf{p}}^{\left( i\right) }\right\rangle ,\hspace{0.1in}%
a_{r}^{(i)}\left| \mathbf{\alpha }_{\mathbf{p}}^{\left( i\right)
}\right\rangle =0\hspace{0.1in}\text{for}\hspace{0.1in}r>0\text{\hspace{%
0.01in}.}
\end{equation}
The irreducible module corresponding to $\left| \mathbf{\alpha }_{\mathbf{p}%
}^{{}}\right\rangle $ is denoted by $H_{\mathbf{p}}:=\bigotimes_{i=1}^{m}H_{%
\mathbf{p}}^{\left( i\right) }$, where $H_{\mathbf{p}}^{\left( i\right) }$
is the irreducible module of $\{a_{k}^{(i)}\}_{k\in \mathbb{Z}}$, defined
by: 
\begin{equation}
H_{\mathbf{p}}^{\left( i\right) }:=\{a_{-n_{q}}^{\left( i\right)
m_{q}}\cdots a_{-n_{1}}^{\left( i\right) m_{1}}\left| \mathbf{\alpha }_{%
\mathbf{p}}^{\left( i\right) }\right\rangle \hspace{0.1in}\text{with}\hspace{%
0.05in}n_{h}>0,\hspace{0.05in}m_{h}>0,\hspace{0.05in}q>0\}.
\end{equation}
Moreover, the vector $\left| \mathbf{\alpha }_{\mathbf{p}}^{\left( i\right)
}\right\rangle $ and the module $H_{\mathbf{p}}^{\left( i\right) }$ are a
h.w. vector and the corresponding irreducible module with respect to the $%
c=1 $ Virasoro algebra $\{L_{n}^{\left( i\right) }\}_{n\in \mathbb{Z}}$
generated by the Heisenberg algebra $\{a_{k}^{(i)}\}_{k\in \mathbb{Z}}$ (see
equation (\ref{Virasoro-Heisenberg})), i.e. 
\begin{equation}
L_{0}^{\left( i\right) }\left| \mathbf{\alpha }_{\mathbf{p}}^{\left(
i\right) }\right\rangle =h_{\mathbf{p}}^{\left( i\right) }\left| \mathbf{%
\alpha }_{\mathbf{p}}^{\left( i\right) }\right\rangle \hspace{0.05in}\text{%
with }h_{\mathbf{p}}^{\left( i\right) }=\frac{1}{2}\mathbf{\alpha }_{\mathbf{%
p}}^{\left( i\right) 2},\hspace{0.05in}\hspace{0.1in}L_{n}^{\left( i\right)
}\left| \mathbf{\alpha }_{\mathbf{p}}^{\left( i\right) }\right\rangle =0%
\hspace{0.05in}\hspace{0.1in}\text{for}\hspace{0.1in}n>0\text{.}
\end{equation}
The zero mode of $m$ independent $c=1$ Virasoro algebras $\{L_{n}^{\left(
i\right) }\}_{n\in \mathbb{Z}}$ is defined by: 
\begin{equation}
L_{0}:=\sum_{i=1}^{m}L_{0}^{\left( i\right) },
\end{equation}
and so we have: 
\begin{equation}
L_{0}\left| \mathbf{\alpha }_{\mathbf{p}}^{{}}\right\rangle =h_{\mathbf{p}%
}\left| \mathbf{\alpha }_{\mathbf{p}}^{{}}\right\rangle
\end{equation}
with 
\begin{equation}
h_{\mathbf{p}}=\sum_{i=1}^{m}h_{\mathbf{p}}^{\left( i\right) }=\mathbf{%
\alpha }_{\mathbf{p}}\mathbf{\alpha }_{\mathbf{p}}^{T}/2.
\end{equation}
The corresponding character is: 
\begin{equation}
Tr_{H_{\mathbf{p}}}\left( q^{\left( L_{0}-\frac{m}{24}\right) }e^{2\pi
iwJ}\right) =\frac{1}{\eta \left( \tau \right) ^{m}}q^{h_{\mathbf{p}%
}}e^{2\pi iw\sqrt{2mp+1}\mathbf{\alpha }_{\mathbf{p}}\mathbf{t}}\text{,}
\end{equation}
where $J:=\mathbf{a}_{0}\mathbf{t}\det (\mathbf{R}_{m,p})$ is the conformal
charge and by definition $H_{\mathbf{p}}$ is the eigenspace of $J$
corresponding to the eigenvalue $\sqrt{2mp+1}\mathbf{\alpha }_{\mathbf{p}}%
\mathbf{t}$.

As in the case of the single free boson CFT (see appendix B), we define the
chiral algebra $\mathfrak{A}(\widehat{u(1)}_{\mathbf{K}_{m,p}})$ extension
of $\mathfrak{A}(\widehat{u(1)})^{\otimes m}$ by adding to it the modes of
the two chiral currents: 
\begin{equation}
\Gamma _{\mathbf{K}_{m,p}}^{\pm }(z):=\hspace{0.03in}:e^{\pm i\mathbf{t}^{T}%
\mathbf{R}_{m,p}\phi \left( z\right) }:,
\end{equation}
where $\phi \left( z\right) ^{T}:=(\phi ^{(1)}(z),...,\phi ^{(m)}(z))$ is
the chiral part of $\mathbf{\varphi }(z,\bar{z})^{T}$. By definition: 
\begin{equation}
\Gamma _{\mathbf{K}_{m,p}}^{\pm }(z)=\bigotimes_{i=1}^{m}\Gamma
_{(2mp+1)}^{\left( i\right) \pm }(z)
\end{equation}
where: 
\begin{equation}
\Gamma _{(2mp+1)}^{\left( i\right) \pm }(z):=:e^{\pm i\sqrt{2mp+1}\phi
^{(i)}(z)}:
\end{equation}
are locally anticommuting Fermi fields, with half integer $(2mp+1)/2$
conformal dimensions.

Let $\left\{ k_{i}\right\} _{i\in \left\{ 1,..,a\right\} }\in \mathbb{Z}_{+}$
and $\left\{ \mathbf{u}_{i,j}\right\} _{j\in \left\{ 1,..,d_{i}\right\} ,%
\text{ }i\in \left\{ 1,..,a\right\} }\in \mathbb{Z}^{m}$ be the eigenvalues
and a basis in $\mathbb{Z}^{m}$ of the corresponding eigenvectors of $%
\mathbf{K}$, $\mathbf{K\bar{u}}_{i,j}=k_{i}\mathbf{u}_{i,j}$, respectively.
Then: 
\begin{equation}
\mathbf{K}\mathbb{Z}^{m}:=\{\mathbf{p}\in \mathbb{Z}^{m}:\mathbf{p=}%
\sum_{i=1}^{a}\sum_{j=1}^{d_{i}}c_{i,j}\mathbf{u}_{i,j}\text{ with }%
c_{i,j}\in \mathbb{Z}\}
\end{equation}
and the quotient $\mathbb{Z}_{\mathbf{K}}:=\mathbb{Z}^{m}/\mathbf{K}\mathbb{Z%
}^{m}$ is: 
\begin{equation}
\mathbb{Z}_{\mathbf{K}}:=\{\mathbf{p}\in \mathbb{Z}^{m}:\mathbf{p=}%
\sum_{i=1}^{a}\sum_{j=1}^{d_{i}}c_{i,j}\mathbf{u}_{i,j}\text{ with }%
c_{i,j}\in \left\{ 0,..,k_{i}-1\right\} \}.
\end{equation}
The matrix $\mathbf{K}_{m,p}$ has two distinct eigenvalues: $k_{1}=2mp+1$,
with degeneracy 1 and eigenvector $\mathbf{t}$, and $k_{2}=1$, with
degeneracy $m-1$ and $m-1$ independent eigenvectors $\left\{ \mathbf{u}%
_{j}\right\} _{j\in \left\{ 1,..,m-1\right\} }\in \mathbb{Z}^{m}$, simply
characterized by the orthogonality condition 
\begin{equation}
\mathbf{u}_{j}^{T}\mathbf{t=0},  \label{ort-condition}
\end{equation}
so that: 
\begin{equation}
\mathbb{Z}_{\mathbf{K}_{m,p}}:=\{\mathbf{p}\in \mathbb{Z}^{m}:\mathbf{p=}b%
\mathbf{t}\text{ with }b\in \left\{ 0,..,2mp\right\} \}.
\end{equation}
Now, $\det (\mathbf{R}_{m,p}^{2})=2mp+1$ and the chiral algebra $\mathfrak{A}%
(\widehat{u(1)}_{\mathbf{K}_{m,p}})$ has h.w. vectors $\left| \mathbf{\alpha 
}_{b\mathbf{t}}\right\rangle $ with conformal weights 
\begin{equation}
\tilde{h}_{b}:=\frac{mb^{2}}{2(2mp+1)}\text{,}
\end{equation}
corresponding to $\mathbf{\alpha }_{b\mathbf{t}}=b\mathbf{t}^{T}\mathbf{R}%
_{m,p}^{-1}$ with $b\mathbf{t}\in \mathbb{Z}_{\mathbf{K}_{m,p}}$. The
related irreducible modules are: 
\begin{equation}
H_{b}:=\bigoplus_{\mathbf{q}\in \mathbb{Z}^{m}}H_{b\mathbf{t+\mathbf{K}}%
_{m,p}\mathbf{q}}\text{,}
\end{equation}
with $b\in \left\{ 0,..,2mp\right\} $, and so the characters are: 
\begin{equation}
\tilde{\chi}_{b}(w|\tau ):=Tr_{H_{b}}\left( q^{\left( L_{0}-\frac{m}{24}%
\right) }e^{2\pi iwJ}\right) \text{\hspace{0.01in}.}
\end{equation}
More explicitly, they are given by: 
\begin{equation}
\tilde{\chi}_{b}(w|\tau )=\frac{1}{\eta \left( \tau \right) ^{m}}%
\sum\limits_{\mathbf{q}\in \mathbb{Z}^{m}}e^{2\pi i\left\{ \frac{\tau }{2}%
\left[ \left( b\mathbf{t}^{T}\mathbf{R}_{m,p}^{-1}+\mathbf{q}^{T}\mathbf{R}%
_{m,p}\right) \left( b\mathbf{t}^{T}\mathbf{R}_{m,p}^{-1}+\mathbf{q}^{T}%
\mathbf{R}_{m,p}\right) ^{T}\right] +w\det (\mathbf{R}_{m,p})\left( b\mathbf{%
t}^{T}\mathbf{R}_{m,p}^{-1}+\mathbf{q}^{T}\mathbf{R}_{m,p}\right) \mathbf{t}%
\right\} }\text{.}
\end{equation}
The chiral algebra $\mathfrak{A}(\widehat{u(1)}_{\mathbf{K}_{m,p}})$ defines
a $\Gamma _{\theta }$-RCFT because its characters $\tilde{\chi}_{b}(w|\tau )$
define a $\left( 2mp+1\right) $-dimensional representation of the modular
subgroup $\Gamma _{\theta }$:

The transformation $T^{2}$: 
\begin{equation}
\tilde{\chi}_{b}(w|\tau +2)=e^{i2\pi \left[ 2\left( \tilde{h}_{b}-\frac{m}{24%
}\right) \right] }\tilde{\chi}_{b}(w|\tau )\text{,}  \label{tra per-per T^2}
\end{equation}
where $\left( \tilde{h}_{b}-m/24\right) $ is the \textit{modular anomaly} of
a h.w. representation of conformal dimension $\tilde{h}_{b}$ in a $\Gamma
_{\theta }$-RCFT with central charge c$\;=m$.

The transformation $S$: 
\begin{equation}
\tilde{\chi}_{b}(\frac{w}{\tau }|-\frac{1}{\tau })=\frac{1}{\sqrt{2pm+1}}%
\sum_{b^{\prime }=0}^{2pm}e^{\frac{2i\pi mbb^{\prime }}{2pm+1}}\tilde{\chi}%
_{b^{\prime }}(w|\tau )\text{\hspace{0.01in}.}
\end{equation}
Such modular transformations can be simply derived from those of the $\Theta 
$-functions with characteristics. We denote this $\Gamma _{\theta }$-RCFT
simply with $\widehat{u(1)}_{\mathbf{K}_{m,p}}$.

The fact that the matrix $\mathbf{K}_{m,p}$ is symmetric implies that $%
\widehat{u(1)}_{\mathbf{K}_{m,p}}$ is invariant under the exchange of a pair
of free bosons. More precisely, the exchange of a pair of free bosons is an
outer automorphism on the chiral algebra $\mathfrak{A}(\widehat{u(1)}_{%
\mathbf{K}_{m,p}})$. Let $g$ be defined as the element that acts on $%
\widehat{u(1)}_{\mathbf{K}_{m,p}}$ bringing the field in position $i$ into
that in position $i+1$, $i\in \left\{ 1,..,m\right\} $, with the periodicity
condition $m+1\equiv 1$. Then, $g$ is an outer automorphism of $\mathfrak{A}(%
\widehat{u(1)}_{\mathbf{K}_{m,p}})$. We observe that $g^{m}\equiv \mathbf{1}$
and $g^{h}\neq \mathbf{1}$ for $h\in \left\{ 1,..,m-1\right\} $, so $g$
generates the discrete symmetry group $\mathbb{Z}_{m}$ of outer
automorphisms of $\mathfrak{A}(\widehat{u(1)}_{\mathbf{K}_{m,p}})$. The
cyclic permutation orbifold in the next section is made with respect to this
discrete symmetry group $\mathbb{Z}_{m}$ of $\widehat{u(1)}_{\mathbf{K}%
_{m,p}}$.\vspace{0.05in}

\begin{prop}
The h.w. representations of \label{clm 1}$\widehat{u(1)}_{\mathbf{K}_{m,p}}$%
\textit{\ can be expressed in terms of those of the tensor product }$%
\widehat{u(1)}_{m\left( 2mp+1\right) }\bigotimes \widehat{su(m)}_{1}$\textit{%
, as the following character decompositions show:} 
\begin{equation}
\tilde{\chi}_{b}(w|\tau )=\sum_{l=0}^{m-1}\chi _{l}^{\widehat{su(m)}%
_{1}}(\tau )K_{(2mp+1)l+mb}^{\left( m\left( 2pm+1\right) \right) }(w|\tau ),
\label{decomp-p-p-ap}
\end{equation}
with\textit{\ }$b\in \left\{ 0,..,2mp\right\} $ \textit{and }$%
K_{b}^{(q)}(w|\tau )$\textit{\ the characters of the }$\Gamma _{\theta }$%
\textit{-RCFT }$\widehat{u(1)}_{q}$\textit{, }$q$\textit{\ odd, given by:} 
\begin{equation}
K_{b}^{\left( q\right) }(w|\tau )=\frac{1}{\eta (\tau )}\Theta \left[ 
\begin{array}{c}
\frac{b}{q} \\ 
0
\end{array}
\right] \left( qw|q\tau \right) \text{.}
\end{equation}
\end{prop}

\textit{Proof ---\ \ }It results: 
\begin{equation}
\mathbf{\alpha }_{b,\mathbf{q}}:=\mathbf{\alpha }_{b\mathbf{t+\mathbf{K}}%
_{m,p}\mathbf{q}}=\sqrt{2pm+1}\left[ \frac{b}{2pm+1}+\left( \frac{l}{m}%
+q\right) \right] \mathbf{t}^{T}+\mathbf{u}^{T}\text{,}  \label{alfa-b-q}
\end{equation}
where $l/m+q:=\left( \sum_{i=1}^{m-1}q_{i}\right) /m$, $l\in \left\{
0,..,m-1\right\} $, $\mathbf{u}:=\mathbf{q-t}\left(
\sum_{i=1}^{m-1}q_{i}\right) /m$ and $\mathbf{q}^{T}:=(q_{1},...,q_{m})\in 
\mathbb{Z}^{m}$. We observe that, denoting $\mathbf{\alpha }_{b,\mathbf{q}%
}:=(\mathbf{\alpha }_{b,\mathbf{q}}^{\left( 1\right) },...,\mathbf{\alpha }%
_{b,\mathbf{q}}^{\left( m\right) })$, it follows: 
\begin{equation}
\mathbf{\alpha }_{b,\mathbf{q}}^{\left( i\right) }-\mathbf{\alpha }_{b,%
\mathbf{q}}^{\left( j\right) }=q_{i}-q_{j}\in \mathbb{Z\hspace{0.08in}}%
\forall i,j\in \left\{ 1,..,m\right\} \text{\hspace{0.01in}.}
\end{equation}
Such a result is at the origin of the decomposition of the modules $H_{b}$
of $\widehat{u(1)}_{\mathbf{K}_{m,p}}$ in terms of the tensor product of
those of the level 1 affine Lie algebra $\widehat{su(m)}_{1}$ and of the
free boson $\widehat{u(1)}_{m\left( 2pm+1\right) }$.

Indeed, to each $\mathbf{\alpha }_{b,\mathbf{q}}$ corresponds the weight $%
\Lambda :=\sum_{i=1}^{m-1}\lambda _{i}\Lambda _{i}$ of $su(m)$, where $%
\Lambda _{i}$ are the fundamental weights of $su(m)$ and $\lambda _{i}$ are
the Dynkin labels, $\lambda _{i}:=q_{i}-q_{i+1}\in \mathbb{Z}$. The $m$%
-ality of the weight $\Lambda $ coincides by definition with $l/m$. We
observe that when $\mathbf{q}$ spans $\mathbb{Z}^{m}$ then $q$ spans $%
\mathbb{Z}$, $l$ spans $\left\{ 0,..,m-1\right\} $ and for any fixed $l$, $%
\Lambda $ spans $\Omega _{l}$.

By using the orthogonality condition $\left( \ref{ort-condition}\right) $
and the definition $\left( \ref{alfa-b-q}\right) $, we obtain: 
\begin{equation}
\mathbf{\alpha }_{b,\mathbf{q}}\mathbf{t}=m\sqrt{2mp+1}\left( \frac{%
mb+\left( 2pm+1\right) l}{m\left( 2pm+1\right) }+q\right) \text{;}
\label{a-b-q}
\end{equation}
so, the conformal dimension $h_{b,\mathbf{q}}:=\mathbf{\alpha }_{b,\mathbf{q}%
}\mathbf{\alpha }_{b,\mathbf{q}}^{T}/2$ has the following expression in
terms of $b$,$\ l$, $q$ and $\Lambda $: 
\begin{equation}
h_{b,\mathbf{q}}=\frac{m\left( 2pm+1\right) }{2}\left( \frac{mb+\left(
2pm+1\right) l}{m\left( 2pm+1\right) }+q\right) ^{2}+h_{\Lambda }\text{,}
\label{h-b-q}
\end{equation}
where $h_{\Lambda }$ is the conformal dimension $\left( \ref{h-lambda}%
\right) $ with $u_{i}:=q_{i}-\left( \sum_{j=1}^{m}q_{j}\right) /m$. The
characters of $\widehat{u(1)}_{\mathbf{K}_{m,p}}$ can then be written in the
form: 
\begin{eqnarray}
\tilde{\chi}_{b}(w|\tau ) &=&\sum_{l=0}^{m-1}\left( \frac{1}{\eta (\tau
)^{m-1}}\sum_{\Lambda \in \Omega _{l}}q^{h_{\Lambda }}\right) \cdot  \notag
\\
&&\cdot \left( \sum\limits_{q\in \mathbb{Z}}\frac{e^{2\pi i\left\{ \tau 
\frac{m\left( 2pm+1\right) }{2}\left( \frac{mb+\left( 2pm+1\right) l}{%
m\left( 2pm+1\right) }+q\right) ^{2}+wm(2mp+1)\left( \frac{mb+\left(
2pm+1\right) l}{m\left( 2pm+1\right) }+q\right) \right\} }}{\eta \left( \tau
\right) }\right)
\end{eqnarray}
which, by $\left( \ref{f-K-P}\right) $ and by the definition of the
characters $K_{a}^{\left( m\left( 2pm+1\right) \right) }(w|\tau )$ of $%
\widehat{u(1)}_{m\left( 2pm+1\right) }$, coincides with \vspace{0.12in}$%
\left( \ref{decomp-p-p-ap}\right) $.

\hfill $\Box $

\vspace{0.15in}

\begin{prop}
\label{clm 2}$\widehat{u(1)}_{\mathbf{K}_{m,p}}$\textit{\ can be seen as a }$%
\Gamma _{\theta }$\textit{-RCFT extension of }$su(m)\bigotimes W_{1+\infty
}^{(m)}\vspace{0.06in}$\textit{, as it follows by the decomposition of its
characters:} 
\begin{equation}
\tilde{\chi}_{b}(w|\tau )=\sum_{\mathbf{q\in }\mathbb{Z}^{\left( m,+\right)
}}d_{su(m)}(\Lambda )\chi _{\mathbf{r(}b\mathbf{,q)}}^{\mathbf{w}%
_{m}}(w|\tau ),  \label{wm-pp}
\end{equation}
\textit{where }$\mathbf{r}(b,\mathbf{q}):=\mathbf{\alpha }_{b,\mathbf{q}}$%
\textit{, }$\mathbb{Z}^{\left( m,+\right) }:=\left\{ \mathbf{q\in }\mathbb{Z}%
^{m}:q_{1}\geq \cdots \geq q_{m}\hspace{0.05in}\right\} $\textit{\ and }$%
\Lambda $ \textit{is defined for }$\mathbf{q\in }\mathbb{Z}^{m}$ as: 
\begin{equation}
\Lambda :=\sum_{i=1}^{m-1}\lambda _{i}\Lambda _{i}\mathbf{\ \ }with\text{: }%
\left\{ 
\begin{array}{c}
\lambda _{i}:=q_{i}-q_{i+1}\in \mathbb{Z\hspace{0.08in}}\forall i\in \left\{
1,..,m-1\right\}  \\ 
l/m+q:=\left( \sum_{i=1}^{m-1}q_{i}\right) /m
\end{array}
\right. ,  \label{landa-q}
\end{equation}
\textit{and }$\Lambda _{i}$\textit{\ the fundamental weights of }$su(m)$%
\textit{.}
\end{prop}

\textit{Proof ---\ \ }The fully degenerate h.w. $\mathbf{r}$, that define
the h.w. module $\mathbb{W}_{\mathbf{r}}^{(m)}$ of $W_{1+\infty }^{(m)}$,
corresponds to the following value of the h.w. $\mathbf{r(}b\mathbf{%
,q):=\alpha }_{b,\mathbf{q}}$ with the only restriction $\mathbf{q\in }%
\mathbb{Z}^{\left( m,+\right) }$. Indeed, by $\left( \ref{alfa-b-q}\right) $ 
$\mathbf{r(}b\mathbf{,q)}$ belongs to $\mathbb{P}^{\left( m\right) }$ if and
only if $\mathbf{q\in }\mathbb{Z}^{\left( m,+\right) }$ and by the
definition $\left( \ref{ch-w-m}\right) $ and the relations $\left( \ref
{h-b-q}\right) $ and $\left( \ref{a-b-q}\right) $ the corresponding
character is: 
\begin{equation}
\chi _{\mathbf{r(}b\mathbf{,q)}}^{\mathbf{w}_{m}}(w|\tau )=\chi _{\Lambda }^{%
\mathcal{W}_{m}}(\tau )\frac{e^{2\pi i\left\{ \tau \frac{m\left(
2pm+1\right) }{2}\left( \frac{mb+\left( 2pm+1\right) l}{m\left( 2pm+1\right) 
}+q\right) ^{2}+wm(2mp+1)\left( \frac{mb+\left( 2pm+1\right) l}{m\left(
2pm+1\right) }+q\right) \right\} }}{\eta \left( \tau \right) }\text{,}
\label{ch-w-(m,p)}
\end{equation}
where $l,$ $q$ and $\Lambda $ are defined by $\left( \ref{landa-q}\right) $.

The identity $\left( \ref{decomp-p-p-ap}\right) $ can be rewritten by using $%
\left( \ref{su(m)-Wm}\right) $ as: 
\begin{equation}
\tilde{\chi}_{b}(w|\tau )=\sum_{l=0}^{m-1}\sum_{\Lambda \in P_{+}\cap \Omega
_{l}}d_{su(m)}(\Lambda )\left( \chi _{\mathbf{\ }\Lambda }^{\mathcal{W}%
_{m}}(\tau )K_{(2mp+1)l+mb}^{\left( m\left( 2pm+1\right) \right) }(w|\tau
)\right) \text{,}  \label{interm-2}
\end{equation}
and so, by $\left( \ref{ch-w-(m,p)}\right) $ and by the definition of the
characters $K_{a}^{\left( m\left( 2pm+1\right) \right) }(w|\tau )$ of $%
\widehat{u(1)}_{m\left( 2pm+1\right) }$, $\left( \ref{interm-2}\right) $
coincides with $\left( \ref{wm-pp}\right) $.

\hfill $\Box $

\vspace{0.15in}

The above identity makes clear the meaning of the claim that $\widehat{u(1)}%
_{\mathbf{K}_{m,p}}$ defines a $\Gamma _{\theta }$-RCFT extension of the
chiral algebra given by the tensor product of the fully degenerate
representations of $W_{1+\infty }^{(m)}$ times the representations of $su(m)$
specialized to $\xi =\left. \left( z\rho \right) \right| _{z=0}$.

In particular, the module $H_{b}$ of $\widehat{u(1)}_{\mathbf{K}_{m,p}}$,
corresponding to the h.w. $\mathbf{\alpha }_{b}:=b\mathbf{t}$, has the
following expansion: 
\begin{equation}
H_{b}=\bigoplus_{\mathbf{q\in }\mathbb{Z}^{\left( m,+\right)
}}F_{su(m)}(\Lambda )\bigotimes \mathbb{W}_{\mathbf{r(}b\mathbf{,q)}}^{(m)}%
\text{,}
\end{equation}
where $F_{su(m)}(\Lambda )$ is the h.w.\ module of $su(m)$ corresponding to
the h.w. $\Lambda $.

\setcounter{equation}{0}

\section{The $\mathbb{Z}_{m}$-orbifold of $\widehat{u(1)}_{\mathbf{K}_{m,p}}$%
}

In this section, we just give the essential elements to identify our TM.
That is, we construct explicitly the $\mathbb{Z}_{m}$ cyclic permutation
orbifold of the $m$-component free bosons $\widehat{u(1)}_{\mathbf{K}_{m,p}}$%
. In particular, a finite set of irreducible characters (modules) of the
orbifold chiral algebra $\mathfrak{A}_{\text{TM}}:=\mathfrak{A}^{\mathbb{Z}%
_{m}}(\widehat{u(1)}_{\mathbf{K}_{m,p}})$ is found. Their modular
transformations have been performed, proving that they give a unitary finite
dimensional representation of the modular subgroup $\Gamma _{\theta }$, i.e.
TM is a $\Gamma _{\theta }$-RCFT.

We refer to our previous paper \cite{Jain} for the construction of the
vertex operators (the chiral primary fields of TM) by the $m$-reduction
procedure \cite{VM}. Furthermore, here we consider the case $m>2$ and prime,
the particular $m=2$ case being developed in \cite{Jain}.

The orbifold construction makes possible to define new RCFTs starting from a
given RCFT by quotienting it with a generic discrete symmetry group $G$. The
discrete group $G$ can be characterized more precisely as a group of
automorphisms of the chiral algebra $\mathfrak{A}$ of the original RCFT. The
orbifold chiral algebra $\mathfrak{A}^{G}:=\mathfrak{A}/G$ is then defined
as the subalgebra of $\mathfrak{A}$ invariant under $G$. The study of the
orbifolds was first introduced in the context of string theory in order to
approximate CFT on Calabi-Yau manifolds \cite{D-H-V-W} and further developed
in \cite{Vafa,D-F-M-S,H-V}. A first detailed study of the general properties
of the orbifolds was done in \cite{DV3}, while in \cite{K-T} a complete
study of orbifolds with respect to discrete groups of inner automorphisms
was given.

Here, we are interested in the class of the cyclic permutation orbifolds. It
was first introduced in \cite{K-S,F-K-S} on RCFTs characterized as the
tensor product of $m$ copies of a given RCFT. The subject was further
developed in \cite{F-S-S,F-R-S,D-M-V-V,B-H-S,Bantay,D-M,B-D-M}, and it was
also studied by using the approach of the $m$-reduction technique in \cite
{VM,cgm,Jain}. A more complete classification has been recently presented in 
\cite{BK}, where the orbifold construction is applied in the more general
framework of the \textit{lattice vertex algebras} (which are not simply
tensor products of RCFTs). In particular, the twisted vertex operator
algebras and their modules are studied.

Let us recall the main steps of the procedure to build the finite set of
irreducible representations of a $\mathbb{Z}_{m}$ cyclic permutation
orbifold. The orbifold chiral algebra $\mathfrak{A}^{\mathbb{Z}_{m}}$ has a
finite set of irreducible representations that splits in two sectors,
untwisted and twisted one. The irreducible representations of the untwisted
sector are generated by restricting those of the original chiral algebra $%
\mathfrak{A}$ to their invariant part with respect to the elements of $%
\mathbb{Z}_{m}$. The characters of the untwisted irreducible $\mathfrak{A}^{%
\mathbb{Z}_{m}}$-representations are not anymore closed under modular
transformations. Then, the irreducible $\mathfrak{A}^{\mathbb{Z}_{m}}$%
-representations of the twisted sector are generated by applying to the
untwisted irreducible $\mathfrak{A}^{\mathbb{Z}_{m}}$-characters the modular
transformations\footnote{%
In our case, the theory to which is applied the orbifold is a $\Gamma
_{\theta }$-RCFT and $m$ is a prime number, so we apply $T^{2j}S\in \Gamma
_{\theta }$ with $j\in \left\{ 0,...,m-1\right\} $.}, $T^{j}S\in PSL(2,%
\mathbb{Z})$ with $j\in \left\{ 0,...,m-1\right\} $.

It is worth pointing out that our theory TM defines a family of lattice
orbifolds which can be included in the general classification presented%
\footnote{%
In particular, the $m$-reduction generates the vertex operators of our
twisted sector, which can be included into the class of twist fields defined
in \cite{BK}.} in \cite{BK}. Indeed, TM describes the cyclic permutation
orbifolds with respect to the outer automorphisms $\mathbb{Z}_{m}$ of the
lattice vertex algebras $\mathfrak{A}(\widehat{u(1)}_{\mathbf{K}_{m,p}})$,
where any $\Gamma _{\theta }$-RCFT $\widehat{u(1)}_{\mathbf{K}_{m,p}}$ is
not a simple tensor product of $m$ copies of a $\Gamma _{\theta }$%
-RCFT.\medskip

\begin{prop}
\label{clm 3}\textit{The theory }TM,\textit{\ characterized as\ the }$%
\mathbb{Z}_{m}$\textit{-orbifold of the }$\Gamma _{\theta }$\textit{-RCFT }$m
$\textit{-component free bosons }$\widehat{u(1)}_{\mathbf{K}_{m,p}}$\textit{%
, has the following content:}

\textbf{The untwisted sector}

\textit{The so called ``P-P'' untwisted sector of }TM\textit{\ coincides
with the }$m$\textit{-component free bosons }$\widehat{u(1)}_{\mathbf{K}%
_{m,p}}$;\textit{\ so, it has }$2pm+1$\textit{\ h.w. representations with
conformal dimensions: } 
\begin{equation}
\tilde{h}_{(b\mathbf{,(1,1))}}=\frac{mb^{2}}{2(2pm+1)}=\,\tilde{h}_{b}\text{,%
}
\end{equation}
\textit{where}$\,\ b=0,..,2mp$\textit{\ and\ corresponding characters:\ } 
\begin{equation}
\tilde{\chi}_{(b\mathbf{,(1,1))}}(w|\tau )=\tilde{\chi}_{b}(w|\tau )\text{.}
\label{ch-M}
\end{equation}

\textit{The so called ``P-A''\ untwisted sector of }TM\textit{\ has }$2pm+1$%
\textit{\ h.w. representations, each one with degeneracy }$m-1$\textit{\ and
conformal dimension: } 
\begin{equation}
\tilde{h}_{(b\mathbf{,(1,}g^{i}\mathbf{))}}=\frac{mb^{2}}{2(2pm+1)}\text{,}
\end{equation}
\textit{where}$\,\ b=0,..,2mp$\textit{, }$i=1,..,m-1$\textit{\ and }$g$ 
\textit{is the generator of the discrete group }$\mathbb{Z}_{m}$\textit{.
The corresponding characters are: } 
\begin{equation}
\tilde{\chi}_{(b\mathbf{,(1,}g^{i}\mathbf{))}}(w|\tau
)=K_{b}^{(2pm+1)}(mw|m\tau )\text{\hspace{0.01in}.}
\label{carattere nontwistato}
\end{equation}

\textbf{The twisted sector}

\textit{The so called ``A-P'' twisted sector of }TM\textit{\ has }$m(2pm+1)$%
\textit{\ h.w.\ representations each one with degeneracy }$m-1$\textit{\ and
conformal dimension: } 
\begin{equation}
h_{(s,f,i)}=\frac{s^{2}}{2(2pm+1)m}+\frac{m^{2}-1}{24m}+\frac{f}{2m}\text{,}
\end{equation}
\textit{where}$\ s=0,..,2mp$\textit{, }$f=0,..,m-1$\textit{\ and }$i=1,..,m-1
$\textit{. Here, the term }$\left( m^{2}-1\right) /24m$\textit{\ takes into
account the conformal dimension of the twist. The characters are: } 
\begin{equation}
\chi _{(s,f,i)}(w|\tau )=\frac{1}{m}\sum_{j=0}^{m-1}e^{-2\pi i\left(
2j\right) (h_{(s,f,i)}-\frac{m}{24})}K_{s}^{\left( 2pm+1\right) }(w|\frac{%
\tau +2j}{m})\text{ .}  \label{carattere twistato}
\end{equation}
\vspace{0.1in}\vspace{0.1in}
\end{prop}

\textit{Proof ---\ \ }Here, we construct all the sectors of the $\mathbb{Z}%
_{m}$-orbifold of $\widehat{u(1)}_{\mathbf{K}_{m,p}}$ showing that they
coincide with those written above in Proposition \ref{clm 3}.

\textbf{The untwisted sector} of the orbifold is obtained introducing a new
h.w. vector $\left| \mathbf{\alpha }_{b\mathbf{t}},\left( \mathbf{1,}\pi
\right) \right\rangle $ and module $H_{b}^{\left( \mathbf{1,}\pi \right) }$
for any element $\pi \in $ $\mathbb{Z}_{m}$ and any h.w. vector\footnote{%
We observe that the h.w. vectors $\left| \mathbf{\alpha }_{b\mathbf{t}%
}\right\rangle $ of $\mathfrak{A}(\widehat{u(1)}_{\mathbf{K}_{m,p}})$ are
invariant under the action of $g$.} $\left| \mathbf{\alpha }_{b\mathbf{t}%
}\right\rangle $ and module $H_{b}$ in the native theory $\widehat{u(1)}_{%
\mathbf{K}_{m,p}}$. The new module $H_{b}^{\left( \mathbf{1,}\pi \right) }$
is defined selecting out from the module $H_{b}$ only the vectors that are
invariant under the action of $\pi $.

We observe that by the definition of $g$ a vector $\left| \mathbf{\alpha }%
^{{}}\right\rangle :=\bigotimes_{i=1}^{m}\left| \mathbf{\alpha }^{\left(
i\right) }\right\rangle $ is invariant under the action of $\pi =g^{i}\,$, $%
i\in \left\{ 1,..,m-1\right\} $, if and only if it is a diagonal vector, $%
\mathbf{\alpha =}a$ $\mathbf{t}^{T}$.

Thus, the irreducible modules $H_{b\mathbf{t+\mathbf{K}}_{m,p}\mathbf{q}}$
of $\mathfrak{A}(\widehat{u(1)})^{\otimes m}$, which can participate to
build $H_{b}^{\left( \mathbf{1,}g^{i}\right) }$ $i\in \left\{
1,..,m-1\right\} $, are only those with diagonal h.w. vector $\left| \mathbf{%
\alpha }_{b\mathbf{t+\mathbf{K}}_{m,p}\mathbf{q}}\right\rangle $. By $\left( 
\ref{alfa-b-q}\right) $, $\left| \mathbf{\alpha }_{b\mathbf{t+\mathbf{K}}%
_{m,p}\mathbf{q}}\right\rangle $ is diagonal if and only if $\mathbf{q}$ is
diagonal, $\mathbf{q:=}q$ $\mathbf{t}^{T}$. So, the vectors in the module $%
H_{b}^{\left( \mathbf{1,}g^{i}\right) }$, $i\in \left\{ 1,..,m-1\right\} $,
are only the diagonal vectors in $H_{b\mathbf{t+\mathbf{K}}_{m,p}q\mathbf{t}%
^{T}}$ with $q\in \mathbb{Z}$.

Summarizing, the modules $H_{b}^{\left( \mathbf{1,}\pi \right) }$ of the
untwisted sector of the $\mathbb{Z}_{m}$-orbifold of $\widehat{u(1)}_{%
\mathbf{K}_{m,p}}$ are defined in the following way: 
\begin{equation}
H_{b}^{\left( \mathbf{1,}\pi \right) }:=\left\{ 
\begin{array}{c}
For\text{ }\pi =g^{0}=\mathbf{1},the\text{ }identity: \\ 
For\,\pi =g^{i}\,\forall i\in \left\{ 1,..,m-1\right\} \,:
\end{array}
\begin{array}{c}
\bigoplus_{\mathbf{q\in }\mathbb{Z}^{m}}H_{b\mathbf{t+\mathbf{K}}_{m,p}%
\mathbf{q}}=H_{b} \\ 
\bigoplus_{q\mathbf{\in }\mathbb{Z}}H_{b,q}^{\left( g\right) }
\end{array}
\right. \text{,}
\end{equation}
where $H_{b,q}^{\left( g\right) }$ is the submodule of $H_{b\mathbf{t+%
\mathbf{K}}_{m,p}q\mathbf{t}^{T}}$ with only diagonal vectors.

The conformal weight of $\left| \mathbf{\alpha }_{b\mathbf{t}},\left( 
\mathbf{1,}\pi \right) \right\rangle $ is: 
\begin{equation}
\tilde{h}_{(b,\left( \mathbf{1,}\pi \right) )}=\frac{mb^{2}}{2(2pm+1)}\,
\label{dim nontwist orb}
\end{equation}
\thinspace \thinspace and the corresponding character is: 
\begin{equation}
\tilde{\chi}_{(b,\left( \mathbf{1,}\pi \right) )}(w|\tau
):=Tr_{H_{b}^{\left( \mathbf{1,}\pi \right) }}\left( q^{\left( L_{0}-\frac{m%
}{24}\right) }e^{2\pi iwJ}\right) \text{.}  \label{carattere nontwistato orb}
\end{equation}
For $\pi =g^{0}=\mathbf{1}$ (the Identity), it reads as: 
\begin{equation}
\tilde{\chi}_{(b,\left( \mathbf{1,1}\right) )}(w|\tau ):=Tr_{H_{b}}\left(
q^{\left( L_{0}-\frac{m}{24}\right) }e^{2\pi iwJ}\right) =\tilde{\chi}%
_{b}(w|\tau )\text{,}  \label{carattere nontwistato orb1}
\end{equation}
while for\thinspace $\ \pi =g^{i},\,$ $i\in \left\{ 1,..,m-1\right\} ,$%
\thinspace it\ reads as: 
\begin{equation}
\tilde{\chi}_{(b,\left( \mathbf{1,}\pi \right) )}(w|\tau
):=Tr_{H_{b}^{\left( \mathbf{1,}\pi \right) }}\left( q^{\left( L_{0}-\frac{m%
}{24}\right) }e^{2\pi iwJ}\right) :=\sum_{q\in \mathbb{Z}}Tr_{H_{b,q}^{%
\left( g\right) }}\left( q^{\left( L_{0}-\frac{m}{24}\right) }e^{2\pi
iwJ}\right) \text{.}
\end{equation}
The definition of $H_{b,q}^{\left( g\right) }$ implies: 
\begin{equation}
Tr_{H_{b,q}^{\left( g\right) }}\left( q^{\left( L_{0}-\frac{m}{24}\right)
}e^{2\pi iwJ}\right) =\frac{1}{\eta \left( m\tau \right) }e^{2\pi i\left\{ 
\frac{\tau }{2}\left[ \left( \mathbf{\alpha }_{b}+q\mathbf{t}^{T}\mathbf{R}%
_{m,p}\right) \left( \mathbf{\alpha }_{b}+q\mathbf{t}^{T}\mathbf{R}%
_{m,p}\right) ^{T}\right] +w\det (\mathbf{R}_{m,p})\left( \mathbf{\alpha }%
_{b}+q\mathbf{t}^{T}\mathbf{R}_{m,p}\right) \mathbf{t}\right\} }
\end{equation}
or more explicitly using $\left( \ref{alfa-b-q}\right) $, $\left( \ref{h-b-q}%
\right) $ and $\left( \ref{a-b-q}\right) $: 
\begin{equation}
Tr_{H_{b,q}^{\left( g\right) }}\left( q^{\left( L_{0}-\frac{m}{24}\right)
}e^{2\pi iwJ}\right) =\frac{1}{\eta \left( m\tau \right) }e^{2\pi i\left\{
\tau \frac{m\left( 2pm+1\right) }{2}\left( \frac{b}{2pm+1}+q\right)
^{2}+wm(2mp+1)\left( \frac{b}{2pm+1}+q\right) \right\} },
\end{equation}
and finally: 
\begin{equation}
\tilde{\chi}_{(b,\left( \mathbf{1,}\pi \right) )}(w|\tau )=K_{b}^{\left(
2pm+1\right) }(mw|m\tau ).  \label{carattere nontwistato orb2}
\end{equation}

The identity $(\ref{carattere nontwistato orb1})$ implies that the h.w. $(%
\mathbf{\alpha }_{b},\left( \mathbf{1,1}\right) )$ representation of the $%
\mathbb{Z}_{m}$-orbifold of $\widehat{u(1)}_{\mathbf{K}_{m,p}}$coincides
with the h.w. $\mathbf{\alpha }_{b}$ representation of $\widehat{u(1)}_{%
\mathbf{K}_{m,p}}$, that is the ``P-P''\ sector coincides with $\widehat{u(1)%
}_{\mathbf{K}_{m,p}}$, while the identity $(\ref{carattere nontwistato orb2}%
) $ implies that there is a $m-1$ degeneracy in the other h.w.
representations that define the ``P-A''\ sector.\vspace{0.1in}

\textbf{The twisted sector} of the orbifold is generated by the action of
the group $\Gamma _{\theta }$ on the untwisted sector. In particular, being $%
\widehat{u(1)}_{\mathbf{K}_{m,p}}$ a $\Gamma _{\theta }$-RCFT the twisted
sector is generated by the action of the group $\Gamma _{\theta }$ on the
``P-A'' untwisted sector.

More precisely, by means of the modular transformation $S$ from the
characters $\tilde{\chi}_{(b,\left( \mathbf{1,}g^{i}\right) )}(w|\tau )$ of
the ``P-A'' untwisted sector we can generate the characters $\tilde{\chi}%
_{(b,\left( g^{i},\mathbf{1}\right) )}(w|\tau )=K_{b}^{\left( 2pm+1\right)
}(w|\frac{\tau }{m}),$ $i\in \left\{ 1,..,m-1\right\} ,$ of the twisted
sector.

Then, using the modular transformation $T^{2}$ on the characters $\tilde{\chi%
}_{(b,\left( g^{i},1\right) )}(w|\tau )$, the following basis in the twisted
sector is obtained: 
\begin{equation}
\tilde{\chi}_{(b,\left( g^{i},g^{2j}\right) )}(w|\tau )=K_{b}^{\left(
2pm+1\right) }(w|\frac{\tau +2j}{m})\text{,}  \label{basis-twist}
\end{equation}
where $b\in \left\{ 0,..,2pm\right\} $, $i\,{\in }\left\{ 1,..,m-1\right\} ${%
\ and }$j\in \left\{ 0,..,m-1\right\} $. Also these characters are
degenerate with respect to the index $i\in \left\{ 1,..,m-1\right\} $.

The invertibility of the equality in $(\ref{carattere twistato})$ implies
that the characters in $(\ref{carattere twistato})$ and $(\ref{basis-twist})$
simply define two different basis of the same twisted sector.\vspace{0.1in}

The reason why we have chosen the characters $\chi _{(s,f,i)}(w|\tau )$ as a
basis is due to the fact that they correspond to well defined h.w.\
representations, as it can be seen by looking at the transformations of
these characters under the elements of the modular subgroup $\Gamma _{\theta
}.$

\hfill $\Box$

\begin{prop}
\label{clm 4}\textit{The theory }TM\textit{\ is a }$\Gamma _{\theta }$%
\textit{-RCFT.\vspace{0.1in}}
\end{prop}

\textit{Proof ---\ \ }We have to show that the characters of TM give a
finite dimensional representation of the modular group $\Gamma _{\theta }$.%
\vspace{0.1in}

The $m$-component free bosons $\widehat{u(1)}_{\mathbf{K}_{m,p}}$ is a $%
\Gamma _{\theta }$-RCFT. The modular transformations of the corresponding
characters are given in section 4.\vspace{0.1in}

The modular transformations of the ``P-A'' untwisted and the ``A-P'' twisted
characters are derived using their definitions $(\ref{carattere nontwistato}%
) $, $(\ref{carattere twistato})$ and the known modular transformations for
the free boson $\Gamma _{\theta }$-RCFT $\widehat{u(1)}_{2mp+1}$, given in
appendix B. \vspace{0.1in}

\textbf{The modular transformations of the characters of the ``P-A''
untwisted sector}

The transformation $T^{2}$: 
\begin{equation}
\tilde{\chi}_{(b\mathbf{,(1,}g^{i}\mathbf{))}}(w|\tau +2)=e^{i2\pi \left[
2\left( \tilde{h}_{(b\mathbf{,(1,}g^{i}\mathbf{))}}-\frac{m}{24}\right) %
\right] }\tilde{\chi}_{(b\mathbf{,(1,}g^{i}\mathbf{))}}(w|\tau )\text{,}
\end{equation}
where $\left( \tilde{h}_{(b\mathbf{,(1,}g^{i}\mathbf{))}}-m/24\right) $ is
the \textit{modular anomaly} of a h.w. representation of conformal dimension 
$\tilde{h}_{(b\mathbf{,(1,}g^{i}\mathbf{))}}$ in a $\Gamma _{\theta }$-RCFT
with central charge c$\;=m$.

The transformation $S$: 
\begin{equation}
\tilde{\chi}_{(b\mathbf{,(1,}g^{i}\mathbf{))}}(\frac{w}{\tau }|-\frac{1}{%
\tau })=\sum_{\mu =0}^{2pm}\sum_{f=0}^{m-1}\frac{e^{\frac{2i\pi b\mu }{2mp+1}%
}}{\sqrt{2mp+1}}\chi _{(\mu ,f,i)}(w|\tau )\text{,}
\end{equation}
it brings the characters of the ``P-A'' untwisted sector $\tilde{\chi}_{(b%
\mathbf{,(1,}g^{i}\mathbf{))}}$ into those of the ``A-P'' twisted sector $%
\chi _{(\mu ,f,i)}$.\vspace{0.1in}

\textbf{The modular transformations of the characters of the ``A-P'' twisted
sector}

The transformation $T^{2}$: 
\begin{equation}
\chi _{(s,f,i)}(w|\tau +2)=e^{i2\pi \left[ 2\left( \tilde{h}_{(s,f,i)}-\frac{%
m}{24}\right) \right] }\chi _{(s,f,i)}(w|\tau )\text{,}
\end{equation}
where $\left( \tilde{h}_{(s,f,i)}-\frac{m}{24}\right) $ is the\textit{\
modular anomaly} of a h.w. representation of conformal dimension $\tilde{h}%
_{(s,f,i)}$ in a $\Gamma _{\theta }$-RCFT with central charge c$\;=m$. 
\vspace{0.1in}

The transformation $S$ on the characters $\chi _{(s,f,i)}$ is the most
subtle to find. By the definition $(\ref{carattere twistato})$ of the
characters $\chi _{(s,f,i)}(w|\tau )$ it is clear that to find their
transformation under $S$ we have to find the action of $S$ on the characters 
$K_{s}^{\left( 2pm+1\right) }(w|\left( \tau +2j\right) /m)$ of $\widehat{u(1)%
}_{2mp+1}$. \vspace{0.1in}

\begin{lemma}
\label{lem 5}\textit{The modular transformation }$S((w|\tau )):=(w/\tau
|-1/\tau )$ \textit{acts on the characters }$K_{s}^{\left( 2pm+1\right)
}(w|(\tau +2j)/m)$\textit{\ of the free boson }$\Gamma _{\theta }$\textit{%
-RCFT }$\widehat{u(1)}_{2mp+1}$\textit{, in the following way:} 
\begin{equation}
S(K_{s}^{\left( 2pm+1\right) }(w|\frac{\tau }{m}))=\sum_{b=0}^{2pm}\left( 
\frac{e^{\frac{2i\pi sb}{2mp+1}}}{\sqrt{2mp+1}}\right) K_{b}^{\left(
2pm+1\right) }(mw|m\tau )  \label{s-0-twist}
\end{equation}
\textit{and} 
\begin{equation}
S(K_{s}^{\left( 2pm+1\right) }(w|\frac{\tau +2j}{m}))=\sum_{b=0}^{2pm}\left( 
\mathcal{A}_{(m,p,2j)}\right) _{(s,b)}K_{b}^{\left( 2pm+1\right) }(w|\frac{%
\tau +2j^{\ast }}{m})\hspace{0.15in}\forall j\in (1,..,m-1)\text{,}
\label{s-j-twist}
\end{equation}
\textit{where }$\left( \mathcal{A}_{(m,p,2j)}\right) _{(s,\mu )}$ \textit{%
are the entries of the }$\left( 2mp+1\right) \times \left( 2mp+1\right) $ 
\textit{matrix }$\mathcal{A}_{(m,p,2j)}$ \textit{that represents the action
of} \textit{the modular transformation }$A_{(m,2j)}\in \Gamma _{\theta }$ 
\textit{on the characters of }$\widehat{u(1)}_{2mp+1}$\textit{. The matrix }$%
A_{(m,2j)}\in \Gamma _{\theta }$\textit{\ and the integer number }$j^{\ast
}\in (1,..,m-1)$\textit{\ are defined in a univocal way by the conditions: } 
\begin{equation}
S(w|\frac{\tau +2j}{m}):=\left( \frac{w}{\tau }|\frac{\left( -1/\tau \right)
+2j}{m}\right) =\ A_{(m,2j)}\left( (w|\frac{\tau +2j^{\ast }}{m})\right) 
\text{\hspace{0.01in},}  \label{eq.AVII14}
\end{equation}
\textit{for any fixed }$j\in (1,..,m-1)$\textit{.}\vspace{0.1in}
\end{lemma}

\textit{Proof of Lemma \ref{lem 5} ---} \ Relation $(\ref{s-0-twist})$ can
be derived in the following way: 
\begin{equation}
S((w|\frac{\tau }{m})):=(\frac{w}{\tau }|-\frac{1}{m\tau })
\end{equation}
but now the right hand side can be seen as the transformation $S^{\prime }$
on the new variables $(w^{\prime }:=mw|\tau ^{\prime }:=m\tau )$, that is $%
S^{\prime }((w^{\prime }|\tau ^{\prime })):=(w^{\prime }/\tau ^{\prime
}|-1/\tau ^{\prime })=(w/\tau |-1/m\tau )$.

Thus, one obtains: 
\begin{equation}
S(K_{s}^{\left( 2pm+1\right) }(w|\frac{\tau }{m}))=S^{\prime }(K_{s}^{\left(
2pm+1\right) }(w^{\prime }|\tau ^{\prime }))
\end{equation}
and, after using the modular transformation $S$ of $K_{s}^{\left(
2pm+1\right) }$ given in appendix B, the equation $(\ref{s-0-twist})$ is
reproduced.

To prove $(\ref{s-j-twist})$ we have to show that the 2$\times $2 matrix $%
A_{(m,2j)}\in \Gamma _{\theta }$ and the integer number $j^{\ast }\in
(1,..,m-1)$ exist and are unique, for any $j$ fixed in $1,..,m-1$.

Indeed, given the matrix $A_{(m,2j)}\in \Gamma _{\theta }$ its
representation $\mathcal{A}_{(m,p,2j)}$ on the characters $K_{s}^{\left(
2pm+1\right) }(w|\tau )$ of the free boson $\Gamma _{\theta }$-RCFT $%
\widehat{u(1)}_{2mp+1}$ follows by the modular transformations given in
appendix B.

By definition, the generic 2$\times $2 matrix $A=\left( 
\begin{array}{cc}
\begin{array}{c}
p \\ 
r
\end{array}
& 
\begin{array}{c}
q \\ 
s
\end{array}
\end{array}
\right) \in PSL(2,\mathbb{Z})$ acts in the following way on $(w|\tau )$: 
\begin{equation}
A\left( (w|\tau )\right) :=\left( \frac{w}{r\tau +s}\left| \frac{p\tau +q}{%
r\tau +s}\right. \right) \text{\hspace{0.01in}.}
\end{equation}
Thus, expanding the right hand side of $(\ref{eq.AVII14})$ it results: 
\begin{equation}
\left( \frac{w}{\tau }\left| \frac{-1+2j\tau }{m\tau }\right. \right)
=\left( \frac{w}{r\left( \tau +2j^{\ast }\right) /m+s}\left| \frac{p\left(
\tau +2j^{\ast }\right) /m+q}{r\left( \tau +2j^{\ast }\right) /m+s}\right.
\right) \text{,}
\end{equation}
whose solution is: 
\begin{equation}
r=m,\hspace{0.1in}p=2j,\hspace{0.1in}s=-2j^{\ast }\text{,}
\end{equation}
where $j^{\ast }$ and $q$ have to satisfy the equation: 
\begin{equation}
\left( 2j\right) \left( 2j^{\ast }\right) +qm=-1\text{\hspace{0.01in}.}
\label{eq-jstar-q}
\end{equation}
The only thing to prove now is that the integer $j^{\ast }\in (1,..,m-1)$
and the \textbf{odd} integer $q$ exist and are unique, for any fixed $j\in
(1,..,m-1)$.

We observe that the equation: 
\begin{equation}
\left( 2j\right) \left( 2j^{\prime }\right) -bm=1  \label{primo}
\end{equation}
has one and only one solution, with $j^{\prime }$ nonzero\ integer number
with minimal modulo and $b$\textbf{\ odd} integer number, for any fixed $%
j\in (1,..,m-1)$.

Indeed, for the hypothesis $m>2$ prime number and $j\in (1,..,m-1)$,
equation $\left( \ref{primo}\right) $ simply expresses that $m$ and $2j$ are
coprime numbers.

Finally, the integer $\alpha $, such that $j^{\ast }=\alpha m-j^{\prime }\in
(1,..,m-1)$, exists and is unique and, putting $q=b-4j\alpha $ \textbf{odd}
integer, the pair $j^{\ast }$ and $q$ satisfy equation $(\ref{eq-jstar-q})$.

Thus, the matrix $A_{(m,2j)}$ is: 
\begin{equation}
\ A_{(m,2j)}=\left( 
\begin{array}{ccc}
\begin{array}{c}
2j \\ 
m
\end{array}
&  & 
\begin{array}{c}
b-4j\alpha \\ 
-2j^{\ast }
\end{array}
\end{array}
\right) \text{\hspace{0.01in}.}
\end{equation}

The only thing left to prove now is that the matrices $A_{(m,2j)}$ are
elements of $\Gamma _{\theta }$\ for any $j\in (1,..,m-1)$.

We observe that $\det A_{(m,2j)}=-\left[ \left( 2j\right) \left( 2j^{\ast
}\right) +\left( b-4j\alpha \right) m\right] $ is 1 using equation $(\ref
{eq-jstar-q})$, so $A_{(m,2j)}\in PSL(2,\mathbb{Z})$. The fact that $%
A_{(m,2j)}\in \Gamma _{\theta }$ is now a direct consequence of the
characterization of $\Gamma _{\theta }$ given in appendix A.

In particular, by using $\left( \ref{pro-T-S def2}\right) $ this matrix can
be expressed in terms of the matrices $T^{2}$ and $S$ as: 
\begin{equation}
\ A_{(m,2j)}=S_{(a_{1},b_{1})}{\times }S_{(a_{2},b_{2})}{\times }...{\times }%
S_{(a_{u},b_{u})}\text{,}  \label{pro-T-S}
\end{equation}
where $u$\ is an odd positive integer, $(a_{h},b_{h})\in \mathbb{Z}{\times }%
\mathbb{Z}$ $\forall h\in (1,..,u)$ and $S_{(a,b)}:=T^{2a}ST^{2b}$.

\hfill $\Box$

\vspace{0.1in}

The results of Lemma \ref{lem 5} make possible to give the modular
transformation $S$ for the characters $\chi _{(\mu ,f,i)}$ of the ``A-P''
twisted sector, according to: 
\begin{eqnarray}
\chi _{(s,f,i)}(\frac{w}{\tau }|-\frac{1}{\tau })
&=&\sum_{b=0}^{2pm}\sum_{e=0}^{m-1}\left( \frac{1}{m}\sum_{j=1}^{m-1}e^{2\pi
i\left[ \left( 2j^{\ast }\right) \left( \tilde{h}_{(b,e,i)}-\frac{m}{24}%
\right) \right] }\left( \mathcal{A}_{(m,p,2j)}\right) _{(s,\mu )}e^{-2\pi i%
\left[ \left( 2j\right) \left( \tilde{h}_{(s,f,i)}-\frac{m}{24}\right) %
\right] }\right) \cdot  \notag \\
&&\cdot \chi _{(b,e,i)}(w|\tau )+\frac{1}{m}\sum_{b=0}^{2pm}\left( \frac{e^{%
\frac{2i\pi sb}{2mp+1}}}{\sqrt{2mp+1}}\right) \tilde{\chi}_{(b\mathbf{,(1,}%
g^{i}\mathbf{))}}(w|\tau )\text{\hspace{0.01in}.}
\end{eqnarray}
\vspace{0.1in}

The previous form of the transformations $T^{2}$ and $S$ (the generators of $%
\Gamma _{\theta }$) for the characters of TM shows that it is a $\Gamma
_{\theta }$-RCFT, so concluding the proof of Proposition \ref{clm 4}.

\hfill $\Box$

\setcounter{equation}{0}

\section{TM as a $\Gamma _{\protect\theta }$-RCFT extension of the fully
degenerate $W_{1+\infty }^{(m)}$}

In all sectors of TM the residual c $=1$ free boson $\Gamma _{\theta }$-RCFT 
$\widehat{u(1)}_{m(2mp+1)}$ can be selected out. This is well evidenced by
the decomposition of the characters of TM in terms of those of $\widehat{u(1)%
}_{m(2mp+1)}$, which is the subject of the following Proposition. \vspace{%
0.1in}

\begin{prop}
\label{clm 5}\textit{The characters of }TM\textit{\ have the following
decomposition in terms of the characters }$K_{p}^{\left( m\left(
2pm+1\right) \right) }(w|\tau )$\textit{\ of }$\widehat{u(1)}_{m(2mp+1)}$:%
\textit{\vspace{0.1in}}

\textit{For the characters of the ``P-P'' untwisted sector}: 
\begin{equation}
\tilde{\chi}_{b}(w|\tau )=\sum_{l=0}^{m-1}\chi _{l}^{\widehat{su(m)}%
_{1}}(\tau )K_{(2mp+1)l+mb}^{\left( m\left( 2pm+1\right) \right) }(w|\tau )%
\text{,}  \label{decomp-p-p}
\end{equation}
\textit{for }$b\in \left\{ 0,..,2pm\right\} $\textit{.}\vspace{0.1in}

\textit{For the characters of the ``P-A'' untwisted sector}:\vspace{0.1in} 
\begin{equation}
\tilde{\chi}_{(b\mathbf{,(1,}g^{i}\mathbf{))}}(w|\tau )=\frac{\eta (\tau )}{%
\eta (m\tau )}K_{mb}^{\left( m\left( 2pm+1\right) \right) }(w|\tau )
\label{decomp-p-a}
\end{equation}
\textit{where }$b\in \left\{ 0,..,2pm\right\} $\textit{\ and }$i\,\in
\left\{ 1,..,m-1\right\} $\textit{.\vspace{0.1in}}

\textit{For the characters of the ``A-P''\ twisted sector}:\vspace{0.1in} 
\begin{equation}
\chi _{(s,f,i)}(w|\tau )=\sum_{l=0}^{m-1}N_{{\large (}l,f-2ls{\large )}%
}(\tau )K_{(2mp+1)l+s}^{\left( m\left( 2pm+1\right) \right) }(w|\tau )\text{,%
}  \label{decomp-a-p}
\end{equation}
\textit{where: } 
\begin{equation}
\,N_{{\large (}l,f{\large )}}(\tau )=\frac{1}{m}\sum_{j=0}^{m-1}e^{-\frac{%
2\pi i}{m}\left( 2j\right) (\frac{f}{2}-\frac{1}{24}-\frac{l^{2}}{2})}\frac{%
\eta (\tau )}{\eta (\frac{\tau +2j}{m})}\text{,}  \label{neutr-tw}
\end{equation}
\textit{for }$s\in \left\{ 0,..,2pm\right\} $\textit{, }$f\in \left\{
0,..,m-1\right\} $\textit{\ and }$i\,\in \left\{ 1,..,m-1\right\} $\textit{.}
\vspace{0.15in}
\end{prop}

\textit{Proof ---\ \ }The decomposition $\left( \ref{decomp-p-p}\right) $ is
derived in Proposition \ref{clm 1}. The decomposition $\left( \ref
{decomp-p-a}\right) $ is an immediate consequence of the definitions of the
characters $\tilde{\chi}_{(b\mathbf{,(1,}g^{i}\mathbf{))}}(w|\tau )$, $%
K_{p}^{\left( q\right) }(w|\tau )$ and of the $\Theta $-functions with
characteristics.

Indeed: 
\begin{align}
& \left. K_{b}^{\left( 2pm+1\right) }(mw|m\tau ):=\frac{1}{\eta (m\tau )}%
\Theta \left[ 
\begin{array}{c}
\frac{b}{2pm+1} \\ 
0
\end{array}
\right] (m\left( 2pm+1\right) w|m\left( 2pm+1\right) \tau )=\right.  \notag
\\
&  \notag \\
& \left. \frac{\eta (\tau )}{\eta (m\tau )}\left( \frac{1}{\eta (\tau )}%
\Theta \left[ 
\begin{array}{c}
\frac{mb}{m\left( 2pm+1\right) } \\ 
0
\end{array}
\right] (m\left( 2pm+1\right) w|m\left( 2pm+1\right) \tau )\right) :=\frac{%
\eta (\tau )}{\eta (m\tau )}K_{mb}^{\left( m\left( 2pm+1\right) \right)
}(w|\tau )\text{.}\right.
\end{align}

The decomposition $\left( \ref{decomp-a-p}\right) $ follows analogously, by
using the following identity, for the $\Theta $-functions with
characteristics: 
\begin{equation}
\Theta \left[ 
\begin{array}{c}
\frac{\lambda }{q} \\ 
0
\end{array}
\right] \left( qw|q\frac{\tau +2j}{m}\right) =\sum_{l=0}^{m-1}e^{2\pi
i\left( 2j\right) \frac{\left( ql+\lambda \right) ^{2}}{2mq}}\Theta \left[ 
\begin{array}{c}
\frac{ql+\lambda }{qm} \\ 
0
\end{array}
\right] \left( mqw|mq\tau \right) \text{,}
\end{equation}
where $\lambda \in (0,.,q-1)$ and $j\in (0,.,m-1)$.

\hfill $\Box$

\vspace{0.15in}

It is worth noticing that in each sector (``P-P'', ``P-A'' and ``A-P'') of
TM the corresponding cosets with respect to $\widehat{u(1)}_{m(2mp+1)}$
define CFTs with central charge c $=m-1$ whose h.w. representations can be
defined in terms of those of the affine Lie algebra $\widehat{su(m)}_{1}$.
In particular, the characters of the h.w. representations of these cosets
are expressed in terms of those of $\widehat{su(m)}_{1}$ but calculated for
different specializations.

The decomposition $\left( \ref{decomp-p-p}\right) $ shows that the
characters of the coset $\left( \text{``P-P''-TM}\right) /\widehat{u(1)}%
_{m\left( 2mp+1\right) }$ are those of the affine Lie algebra $\widehat{su(m)%
}_{1}\ $specialized at $\hat{\xi}=\left. \left( z\rho +\tau \widehat{\Lambda 
}_{0}\right) \right| _{z=0}$.

The decomposition $\left( \ref{decomp-p-a}\right) $ and the identity $\left( 
\ref{F-2}\right) $ show that the characters of the coset\linebreak $\left( 
\text{``P-A''-TM}\right) /\widehat{u(1)}_{m\left( 2mp+1\right) }$ are those
of the affine Lie algebra $\widehat{su(m)}_{1}\ $specialized at $\hat{\xi}%
=\rho /m+\tau \widehat{\Lambda }_{0}$.

The decomposition $\left( \ref{decomp-a-p}\right) $, $\left( \ref{neutr-tw}%
\right) $ and the identity $\left( \ref{twistato neta ratio}\right) $ show
that the characters of the coset\linebreak $\left( \text{``A-P''-TM}\right) /%
\widehat{u(1)}_{m\left( 2mp+1\right) }$ are written in terms of those of the
affine Lie algebra $\widehat{su(m)}_{1}\ $specialized at $\hat{\xi}=\rho
\tau /m+\tau \widehat{\Lambda }_{0}$ times the function $F_{twist}^{(m)}(%
\tau )$, that account for the twist with conformal dimension $\left(
m^{2}-1\right) /24m$.\vspace{0.05in}

Finally, the above observations together with the results of section \ref
{main results} make possible to show:

\begin{prop}
\label{clm 6}\textit{The theory }TM\textit{\ is a }$\Gamma _{\theta }$%
\textit{-RCFT extension of the fully degenerate }$W_{1+\infty }^{(m)}$%
\textit{, as it follows by the decomposition of its characters in terms of
those of the fully degenerate }$W_{1+\infty }^{(m)}$:\textit{\vspace{0.1in}}

\textit{For the characters\ of the ``P-P'' untwisted sector}:\textbf{\vspace{%
0.1in}} 
\begin{equation}
\tilde{\chi}_{b}(w|\tau )=\sum_{\mathbf{q\in }\mathbb{Z}^{\left( m,+\right)
}}d_{su(m)}(\Lambda )\chi _{\mathbf{r(}b\mathbf{,q)}}^{\mathbf{w}%
_{m}}(w|\tau )\text{,}  \label{wm-pp+}
\end{equation}
\textit{where }$\mathbf{r(}b\mathbf{,q):=}b\mathbf{t}^{T}\mathbf{R}%
_{m,p}^{-1}+\mathbf{qR}_{m,p}$\textit{, with }$b\in \left\{ 0,..,2pm\right\} 
$\textit{.\vspace{0.1in}}

\textit{For the characters of the ``P-A'' untwisted sector}:\textbf{\vspace{%
0.1in}} 
\begin{equation}
\tilde{\chi}_{(b\mathbf{,(1,}g^{i}\mathbf{))}}(w|\tau )=\sum_{\mathbf{q\in }%
\mathbb{Z}^{\left( m,+\right) }\cap D_{m}}\epsilon (w_{\Lambda })\chi _{%
\mathbf{r(}b\mathbf{,q)}}^{\mathbf{w}_{m}}(w|\tau )\text{,}  \label{wm-pa}
\end{equation}
\textit{where} $b\in \left\{ 0,..,2pm\right\} $\textit{\ and }$i\,\in
\left\{ 1,..,m-1\right\} $\textit{.\vspace{0.1in}}

\textit{For the characters of the ``A-P'' twisted sector}:\textbf{\vspace{%
0.1in}} 
\begin{equation}
\chi _{(s,f,i)}(w|\tau )=F_{twist}^{(m)}(\tau
)\sum_{j=0}^{m-1}\sum_{l=0}^{m-1}\sum_{a=0}^{m-1}\{H_{(s,f,l,a,j)}\sum_{%
\mathbf{q\in }\mathbb{Z}_{a}^{\left( m,+\right) }}\chi _{\Lambda }^{su(m)}(%
\frac{\rho }{m}(\tau +2j))\chi _{\mathbf{r(}s,l,\mathbf{q)}}^{\mathbf{w}%
_{m}}(w|\tau )\}\text{,}  \label{wm-ap}
\end{equation}
\textit{where }$\mathbb{Z}_{a}^{\left( m,+\right) }:=\left\{ \mathbf{q\in }%
\mathbb{Z}^{\left( m,+\right) }:\left( \sum_{i=1}^{m-1}q_{i}\right) =a\text{ 
}mod\hspace{0.02in}m\right\} $,\textit{\ }$\mathbf{r(}s,l,\mathbf{q):=}\left[
\left( s+l-a\right) /m\right] \mathbf{t}^{T}\mathbf{R}_{m,p}^{-1}+\mathbf{qR}%
_{m,p}$\textit{\ with }$\mathbf{q\in }\mathbb{Z}_{a}^{\left( m,+\right) }$, 
\begin{equation}
H_{(s,f,l,a,j)}:=\frac{1}{m}\mathbf{S}_{0,a}^{\widehat{su(m)}_{1}}e^{\frac{%
2\pi i}{m}j\left[ 2ls+l^{2}-f+a(m-a)\right] }\text{,}
\end{equation}
$s\in \left\{ 0,..,2pm\right\} $\textit{, }$f\in \left\{ 0,..,m-1\right\} $%
\textit{\ and }$i\,\in \left\{ 1,..,m-1\right\} $\textit{.\vspace{0.1in}}

\textit{In }$\left( \ref{wm-pp+}\right) $, $\left( \ref{wm-pa}\right) $%
\textit{\ and }$\left( \ref{wm-ap}\right) $\textit{\ }$\Lambda $\textit{\ is
always defined by }$\mathbf{q}$\textit{\ according to }$\left( \ref{landa-q}%
\right) $\textit{.\vspace{0.05in}}
\end{prop}

\textit{Proof ---\ \ }Equation $\left( \ref{wm-pp+}\right) $ is the subject
of Proposition \ref{clm 2}. The proof of $\left( \ref{wm-pa}\right) $\textit{%
\ }is a consequence of decomposition $\left( \ref{decomp-p-a}\right) $ and
of the identities $\left( \ref{F-2}\right) $ and $\left( \ref{F-1}\right) $.
Indeed, the same considerations of Proposition \ref{clm 2} imply: 
\begin{equation}
\tilde{\chi}_{(b\mathbf{,(1,}g^{i}\mathbf{))}}(w|\tau )=\sum_{\mathbf{q\in }%
\mathbb{Z}^{\left( m,+\right) }}\chi _{\Lambda }^{su(m)}(\frac{\rho }{m}%
)\chi _{\mathbf{r(}b\mathbf{,q)}}^{\mathbf{w}_{m}}(w|\tau )\text{,}
\end{equation}
that leads to $\left( \ref{wm-pa}\right) $ by $\left( \ref{su(m)-ro/m}%
\right) $.

Finally, the proof of $\left( \ref{wm-ap}\right) $ follows by the
decomposition $\left( \ref{decomp-p-a}\right) $ and the Corollary \ref{cor}.
Indeed, this last one implies: 
\begin{equation}
\frac{\eta (\tau )}{\eta (\frac{\tau +2j}{m})}=F_{twist}^{(m)}(\tau
)\sum_{a=0}^{m-1}e^{2\pi i\left( 2j\right) \left( h_{\widehat{\Lambda }_{a}}-%
\frac{1}{24m}\right) }\mathbf{S}_{0,a}^{\widehat{su(m)}_{1}}\chi _{\widehat{%
\Lambda }_{a}}^{\widehat{su(m)}_{1}}(\frac{\rho }{m}\left( \tau +2j\right)
|\tau )\text{,}
\end{equation}
that becomes by $\left( \ref{coset-1}\right) $: 
\begin{equation}
\frac{\eta (\tau )}{\eta (\frac{\tau +2j}{m})}=F_{twist}^{(m)}(\tau
)\sum_{a=0}^{m-1}e^{2\pi i\left( 2j\right) \left( h_{\widehat{\Lambda }_{a}}-%
\frac{1}{24m}\right) }\mathbf{S}_{0,a}^{\widehat{su(m)}_{1}}\sum_{\Lambda
\in P_{+}\cap \Omega _{a}}\chi _{\Lambda }^{su(m)}(\frac{\rho }{m}\left(
\tau +2j\right) )\chi _{\mathbf{\ }\Lambda }^{\mathcal{W}_{m}}(\tau )\text{.}
\end{equation}
Now, following the same consideration developed in the proof of Proposition 
\ref{clm 2} and using the decomposition $\left( \ref{decomp-a-p}\right) $ it
results: 
\begin{align}
& \left. \chi _{(s,f,i)}(w|\tau )=F_{twist}^{(m)}(\tau
)\sum_{j=0}^{m-1}\sum_{l=0}^{m-1}\sum_{a=0}^{m-1}\right.  \notag \\
& \text{ \ \ \ \ \ \ \ \ }\{\frac{1}{m}\mathbf{S}_{0,a}^{\widehat{su(m)}%
_{1}}e^{2\pi i\left( 2j\right) \left( \frac{\left( \left( 2pm+1\right)
l+s\right) ^{2}}{2m\left( 2pm+1\right) }+h_{\widehat{\Lambda }_{a}}+\frac{%
m^{2}-1}{24m}-h_{(s,f,i)}\right) }\sum_{\mathbf{q\in }\mathbb{Z}_{a}^{\left(
m,+\right) }}\chi _{\Lambda }^{su(m)}(\frac{\rho }{m}(\tau +2j))\chi _{%
\mathbf{r(}s,l,\mathbf{q)}}^{\mathbf{w}_{m}}(w|\tau )\}\text{,}
\label{interm-rel}
\end{align}
where $\mathbf{r}(s,l,\mathbf{q}):=\left[ \left( s+l-a\right) /m\right] 
\mathbf{t}^{T}R_{m,p}^{-1}+\mathbf{q}R_{m,p}$\ with $\mathbf{q}\in
Z_{a}^{\left( m,+\right) }$. Equation (\ref{interm-rel}) coincides with $%
\left( \ref{wm-ap}\right) $ taking into account that: 
\begin{equation}
h_{\widehat{\Lambda }_{a}}:=\frac{\left( \widehat{\Lambda }_{a},\widehat{%
\Lambda }_{a}+2\rho \right) }{2(m+1)}=\frac{a(m-a)}{2m}\text{.}
\end{equation}

\hfill $\Box $\hfill

\section{Conclusions}

In this paper, we found an RCFT extension of the fully degenerate $%
W_{1+\infty }^{(m)}$ chiral algebra. The relevance of the last chiral
algebra for the description of the Quantum Hall Fluid plateaux has been
underlined in \cite{Cappelli}. The TM model has been applied to the
description of such a phenomenon in \cite{cgm,Jain} and to other physical
systems in \cite{cmn}. An interesting property of such an RCFT is the
possibility of defining different extensions of $W_{1+\infty }^{(m)}$ in any
sector of the orbifold. That relies deeply on the different multiplicities
of the physical vectors appearing in the spectrum of each sector. Moreover,
we found that there is a one to one correspondence between the CFTs with
chiral symmetry $su(m)\bigotimes W_{1+\infty }^{(m)}$ \cite{FZ} and the so
called minimal models \cite{Cappelli}. They are simply two different sectors
of TM, which so gives a consistent RCFT containing fully degenerate
representations of $W_{1+\infty }^{(m)}$ and satisfying the modular
invariance constraint (i.e. it is a completion of the minimal model given in 
\cite{Cappelli}).\medskip

\textbf{Acknowledgments.}~~G.N. was supported by a postdoctoral fellowship
of the Minist\`{e}re fran\c{c}ais d\'{e}l\'{e}gu\'{e} \`{a} l'Enseignement
sup\'{e}rieur et \`{a} la Recherche.\newpage \setcounter{equation}{0} 
\appendix

\section{\textbf{The }$\Gamma _{\protect\theta }$\textbf{\ group}\protect%
\vspace{0.15in}}

The group $\Gamma _{\theta }$ denotes, according to the definition given in
section 13.4 of the Kac's book \cite{Kac-book}, the subgroup of the modular
group $PSL(2,\mathbb{Z})$ generated by $T^{2}$ and $S$. A matrix
representation of the generators $T^{2}$ and $S$ is: 
\begin{equation}
T^{2}=\left( 
\begin{array}{ccc}
\begin{array}{c}
1 \\ 
0
\end{array}
&  & 
\begin{array}{c}
2 \\ 
1
\end{array}
\end{array}
\right) \,\,,\,\,S=\left( 
\begin{array}{ccc}
\begin{array}{c}
0 \\ 
1
\end{array}
&  & 
\begin{array}{c}
-1 \\ 
0
\end{array}
\end{array}
\right) \text{.}
\end{equation}
$\vspace{0.1in}\Gamma _{\theta }$ is the group of elements: 
\begin{equation}
\Gamma _{\theta }=\left\{ \left( 
\begin{array}{ccc}
\begin{array}{c}
a \\ 
c
\end{array}
&  & 
\begin{array}{c}
b \\ 
d
\end{array}
\end{array}
\right) \in PSL(2,\mathbb{Z})\,\,:\,\,a+d\,,\,\,b+c\,\,even,\,\,a+b\,\,odd%
\right\} \text{\hspace{0.01in}.}
\end{equation}
Any element $A$ of$\ \Gamma _{\theta }$ can be represented as follows: 
\begin{equation}
A=\left\{ 
\begin{array}{c}
T^{2a}\,\,\,\,\forall \,\,a\in \mathbb{Z} \\ 
\\ 
S_{(a_{1},b_{1})}{\times }S_{(a_{2},b_{2})}{\times }...{\times }%
S_{(a_{r},b_{r})}\,\left. 
\begin{array}{c}
\forall (a_{j},b_{j})\in \mathbb{Z}{\times }\mathbb{Z}\,,\,\forall j\in
(1,..,r) \\ 
\forall r\in \mathbf{N}
\end{array}
\right.
\end{array}
\right.
\end{equation}
where $S_{(a,b)}=T^{2a}ST^{2b}$. Thus, the characterization given for the
subgroup $\Gamma _{\theta }$ is a direct consequence of the form of these
matrices; indeed: 
\begin{equation}
T^{2a}=\left( 
\begin{array}{ccc}
\begin{array}{c}
1 \\ 
0
\end{array}
&  & 
\begin{array}{c}
2a \\ 
1
\end{array}
\end{array}
\right)
\end{equation}
and 
\begin{equation}
\prod\limits_{j=1}^{r}S_{(a_{j},b_{j})}=\left\{ 
\begin{array}{c}
\left( 
\begin{array}{ccc}
\begin{array}{c}
2\alpha +1 \\ 
2\gamma
\end{array}
&  & 
\begin{array}{c}
2\beta \\ 
2\delta +1
\end{array}
\end{array}
\right) \,\,\,for\,\,r\,\,even\, \\ 
\\ 
\,\left( 
\begin{array}{ccc}
\begin{array}{c}
2\alpha \\ 
2\gamma +1
\end{array}
&  & 
\begin{array}{c}
2\beta +1 \\ 
2\delta
\end{array}
\end{array}
\right) \,\,\,for\,\,r\,\,odd\,
\end{array}
\right. \text{,}  \label{pro-T-S def2}
\end{equation}
where $\alpha $, $\beta $, $\gamma $, $\delta $ are integers and depend on $%
(a_{j},b_{j})$ and $r$.\newpage \setcounter{equation}{0}

\section{\textbf{The }$\Gamma _{\protect\theta }$\textbf{-RCFT} $\widehat{%
u(1)}_{p}$}

Let us recall that the quantized free boson field has the following
expansion: 
\begin{equation}
\varphi (z,\bar{z}):=\varphi _{0}+\phi \left( z\right) +\bar{\phi}\left( 
\bar{z}\right) ,
\end{equation}
$\hspace{0.1in}$where: 
\begin{equation}
\text{ }\left\{ 
\begin{array}{c}
\phi \left( z\right) :=ia_{0}\ln \left( 1/z\right) +i\sum_{k\in \mathbb{Z}%
-\left\{ 0\right\} }a_{k}z^{-k}/k\vspace{0.05in} \\ 
\bar{\phi}\left( \bar{z}\right) :=i\bar{a}_{0}\ln \left( 1/\bar{z}\right)
+i\sum_{k\in \mathbb{Z}-\left\{ 0\right\} }\bar{a}_{k}\bar{z}^{-k}/k
\end{array}
\right. \text{,}
\end{equation}
$\left\{ a_{k}\right\} _{k\in \mathbb{Z}}$ and $\left\{ \bar{a}_{k}\right\}
_{k\in \mathbb{Z}}$ are two independent chiral Heisenberg algebra $%
\mathfrak{A}(\widehat{u(1)})$ and the zero-mode $\varphi _{0}$ is a
conjugate operator to $a_{0}$ $\left( \bar{a}_{0}\right) $: 
\begin{equation}
\left[ a_{n},a_{m}\right] =n\delta _{n,m},\hspace{0.05in}\left[ a_{n},\bar{a}%
_{m}\right] =0,\hspace{0.05in}\left[ \bar{a}_{n},\bar{a}_{m}\right] =n\delta
_{n,m},\hspace{0.05in}\left[ \varphi _{0},a_{m}\right] =i\delta _{0,m},%
\hspace{0.05in}\left[ \varphi _{0},\bar{a}_{m}\right] =i\delta _{0,m}\hspace{%
0.04in}.
\end{equation}
The free boson $\widehat{u(1)}$ with chiral algebra $\mathfrak{A}(\widehat{%
u(1)})$, generated by the modes of the conserved current $i\partial \phi
\left( z\right) $, is a chiral CFT with stress energy tensor $T(z):=\left(
-1/2\right) \hspace{0.02in}$:$\partial \phi \left( z\right) \partial \phi
\left( z\right) $: and central charge $c=1$. This CFT has a one parameter
family of h.w. vectors: 
\begin{equation}
a_{0}\left| \alpha \right\rangle =\alpha \left| \alpha \right\rangle ,%
\hspace{0.1in}a_{n}\left| \alpha \right\rangle =0\hspace{0.1in}\text{for}%
\hspace{0.1in}n>0,
\end{equation}
with corresponding (h.w.) irreducible positive energy module $H_{\alpha
}:=\{a_{-n_{q}}^{m_{q}}\cdots a_{-n_{1}}^{m_{1}}\left| \alpha \right\rangle 
\hspace{0.1in}$with$\hspace{0.05in}n_{i}>0,\hspace{0.05in}m_{i}>0,\hspace{%
0.05in}q>0\}$. The module $H_{\alpha }$ is the irreducible module of the
Virasoro algebra with h.w. $\alpha $ and conformal dimension\footnote{%
It is, in fact, the lowest eigenvalue of $L_{0}$ in the module $H_{\alpha }$.%
} $\alpha ^{2}/2$, as the expansions: 
\begin{equation}
L_{n}:=\frac{1}{2}\sum_{m\in \mathbb{Z}}a_{n-m}a_{m}\hspace{0.05in}\hspace{%
0.05in}\forall n\in \mathbb{Z}-\left\{ 0\right\} ,\hspace{0.05in}%
L_{0}:=\sum_{m\in \mathbb{Z}}a_{-m}a_{m}\hspace{0.05in}+\frac{1}{2}a_{0}^{2}
\label{Virasoro-Heisenberg}
\end{equation}
in the modes of the Heisenberg algebra $\mathfrak{A}(\widehat{u(1)})$ imply.

The free boson $\widehat{u(1)}$ of course is not a rational CFT. RCFT
extensions \cite{BMT} of it are defined compactifying the free boson field
on a circle of rational square radius and correspondingly introducing an
extension of the Heisenberg algebra $\mathfrak{A}(\widehat{u(1)})$. More
explicitly, the compactification condition on the circle with radius $r=%
\sqrt{2p}$, $p$ positive integer, is: 
\begin{equation}
\varphi (ze^{2\pi i},\bar{z}e^{-2\pi i})=\varphi (z,\bar{z})+2\pi rm\text{,}
\label{comp-free-bos1}
\end{equation}
where$\,\,m\in \mathbb{Z}$ is called \textit{winding number }\cite{Di
Francesco}. The compactification condition has the only effect to influence
the zero-mode $a_{0}$ $\left( \bar{a}_{0}\right) $ of the free boson field.
In particular, to obtain well defined vertex operators under the
compactification condition, the possible eigenvalues of $a_{0}$ are
restricted to the following values $\alpha _{n}:=n/r$, $\,n\in \mathbb{Z}$.
So, the h.w. vectors of the compactified free boson $\widehat{u(1)}$ get
reduced to: 
\begin{equation}
a_{0}\left| \alpha _{n}\right\rangle =\alpha _{n}\left| \alpha
_{n}\right\rangle ,\hspace{0.1in}a_{r}\left| \alpha _{n}\right\rangle =0%
\hspace{0.1in}\text{for}\hspace{0.1in}r>0,\hspace{0.1in}L_{0}\left| \alpha
_{n}\right\rangle =h_{n}\left| \alpha _{n}\right\rangle \text{,}
\end{equation}
where $h_{n}:=\alpha _{n}^{2}/2$. The irreducible module corresponding to
the h.w. vector $\left| \alpha _{n}\right\rangle $ is denoted by $H_{n}$ and
the corresponding character is: 
\begin{equation}
Tr_{H_{n}}\left( q^{\left( L_{0}-\frac{m}{24}\right) }e^{2\pi iwJ}\right) =%
\frac{1}{\eta \left( \tau \right) }q^{h_{n}}e^{2\pi iwr\alpha _{n}}\text{,}
\end{equation}
where $J:=ra_{0}$ is the conformal charge.

The chiral algebra $\mathfrak{A}(\widehat{u(1)}_{2p})$ extension of the
Heisenberg algebra $\mathfrak{A}(\widehat{u(1)})$ is defined by adding to it
the modes of the two chiral currents $\Gamma _{2p}^{\pm }(z):=\hspace{0.03in}
$:$e^{\pm i\sqrt{2p}\phi \left( z\right) }$:\hspace{0.03in}. $\Gamma
_{2p}^{\pm }(z)$ are uniquely characterized as the vertex operators with
lowest nonzero conformal dimension satisfying the requirements of well
definition with respect to the compactification condition $\left( \ref
{comp-free-bos1}\right) $ (i.e. they have to be invariant under $\phi
\rightarrow \phi +2\pi r$) and of locality (i.e. integer conformal
dimension). The chiral algebra $\mathfrak{A}(\widehat{u(1)}_{2p})$ has $%
r^{2}=2p$ h.w. vectors $\left| \alpha _{l}\right\rangle $, those with $%
\alpha _{l}=l/r$ for $l\in \left\{ 0,.,2p-1\right\} $. The corresponding
irreducible modules are $H_{l}^{(2p)}:=\bigoplus_{u\in \mathbb{Z}%
}H_{l+u\left( 2p\right) }$ and so the corresponding characters are: 
\begin{equation}
K_{l}^{\left( 2p\right) }(w|\tau ):=Tr_{H_{l}^{(2p)}}\left( q^{\left( L_{0}-%
\frac{m}{24}\right) }e^{2\pi iwJ}\right) =\frac{1}{\eta \left( \tau \right) }%
\sum_{u\in \mathbb{Z}}q^{p(\frac{l}{2p}+u)^{2}}e^{2\pi iw\left( 2p\right) (%
\frac{l}{2p}+u)}\text{,}
\end{equation}
for $l\in \left\{ 0,.,2p-1\right\} $, which in terms of $\Theta $-functions
with characteristics are written as: 
\begin{equation}
K_{l}^{\left( 2p\right) }(w|\tau )=\frac{1}{\eta (\tau )}\Theta \left[ 
\begin{array}{c}
\frac{l}{2p} \\ 
0
\end{array}
\right] \left( 2pw|2p\tau \right) \text{\hspace{0.01in}.}
\end{equation}
The chiral algebra $\mathfrak{A}(\widehat{u(1)}_{2p})$ defines an RCFT
because its characters $K_{l}^{\left( 2p\right) }(w|\tau )$ define a $2p$%
-dimensional representation of the entire modular group $PSL(2,\mathbb{Z})$.
That immediately follows by the modular transformations of\ the characters $%
K_{l}^{\left( 2p\right) }(w|\tau )$: 
\begin{equation}
K_{l}^{\left( 2p\right) }(w|\tau +1)\text{{}}=\text{{}}e^{i2\pi \left( \frac{%
l^{2}}{4p}-\frac{1}{24}\right) }K_{l}^{\left( 2p\right) }(w|\tau ),\hspace{%
0.1in}\hspace{0.1in}K_{l}^{\left( 2p\right) }(\frac{w}{\tau }|-\frac{1}{\tau 
})\text{{}}=\text{{}}\frac{1}{\sqrt{2p}}\sum_{l^{\prime }=0}^{2p-1}e^{\frac{%
2i\pi l^{\prime }l}{2p}}K_{l^{\prime }}^{\left( 2p\right) }(w|\tau )\text{%
\hspace{0.01in}.}
\end{equation}
We denote this RCFT simply with $\widehat{u(1)}_{2p}$.

Here, we want to define a class of $\Gamma _{\theta }$-RCFT extensions of
the Heisenberg algebra $\mathfrak{A}(\widehat{u(1)})$. It can be done by
admitting for the free boson CFT a compactification condition with odd
square radius $r^{2}=p$, $p$ odd.

The same analysis as above holds with the only difference that the chiral
algebra $\mathfrak{A}(\widehat{u(1)}_{p})$\ giving the extension of the
Heisenberg algebra $\mathfrak{A}(\widehat{u(1)})$ is now defined by adding
to it the modes of the two chiral currents $\Gamma _{p}^{\pm }(z):=\hspace{%
0.03in}$:$e^{\pm i\sqrt{p}\phi \left( z\right) }$:\hspace{0.03in}, which are
now locally anticommuting Fermi fields with half integer ($p/2$) conformal
dimensions.

The chiral algebra $\mathfrak{A}(\widehat{u(1)}_{p})$ has $r^{2}=p$ h.w.
vectors $\left| \alpha _{l}\right\rangle $, those with $\alpha _{l}=l/r$ for 
$l\in \left\{ 0,.,p-1\right\} $. The corresponding irreducible modules are $%
H_{l}^{(p)}:=\bigoplus_{u\in \mathbb{Z}}H_{l+up}$ and so the corresponding
characters are: 
\begin{equation}
K_{l}^{\left( p\right) }(w|\tau ):=Tr_{H_{l}^{(p)}}\left( q^{\left( L_{0}-%
\frac{m}{24}\right) }e^{2\pi iwJ}\right) =\frac{1}{\eta \left( \tau \right) }%
\sum_{u\in \mathbb{Z}}q^{\frac{p}{2}(\frac{l}{p}+u)^{2}}e^{2\pi iwp(\frac{l}{%
p}+u)}\text{,}
\end{equation}
where $J:=ra_{0}$, $l\in \left\{ 0,.,p-1\right\} $, or in terms of $\Theta $%
-functions: 
\begin{equation}
K_{l}^{\left( p\right) }(w|\tau )=\frac{1}{\eta (\tau )}\Theta \left[ 
\begin{array}{c}
\frac{l}{p} \\ 
0
\end{array}
\right] \left( pw|p\tau \right) \text{\hspace{0.01in}.}
\end{equation}
The chiral algebra $\mathfrak{A}(\widehat{u(1)}_{p})$ defines a $\Gamma
_{\theta }$-RCFT because its characters $K_{l}^{\left( p\right) }(w|\tau )$
define a $p$-dimensional representation of the modular subgroup $\Gamma
_{\theta }$. Indeed, the modular transformation of\ $K_{l}^{\left( p\right)
}(w|\tau )$ are: 
\begin{equation}
K_{l}^{\left( p\right) }(w|\tau +2)\text{{}}=\text{{}}e^{i4\pi \left( \frac{%
l^{2}}{2p}-\frac{1}{24}\right) }K_{l}^{\left( p\right) }(w|\tau ),\hspace{%
0.1in}\hspace{0.1in}K_{l}^{\left( p\right) }(\frac{w}{\tau }|-\frac{1}{\tau }%
)\text{{}}=\text{{}}\frac{1}{\sqrt{p}}\sum_{l^{\prime }=0}^{p-1}e^{\frac{%
2i\pi l^{\prime }l}{p}}K_{l^{\prime }}^{\left( p\right) }(w|\tau )\text{%
\hspace{0.01in}.}
\end{equation}
We denote this $\Gamma _{\theta }$-RCFT simply as $\widehat{u(1)}_{p}$, $p$
odd.

Furthermore, we observe that the $\Gamma _{\theta }$-RCFT $\widehat{u(1)}%
_{p} $ , $p$ odd, coincides with the $\Gamma _{\theta }$-projection of the
ordinary RCFT $\widehat{u(1)}_{4p}$. This is an immediate consequence of the
relations among the corresponding characters: 
\begin{equation}
K_{l}^{\left( p\right) }(w|\tau )=K_{2l}^{\left( 4p\right) }(w|\tau
)+K_{2(p+l)}^{\left( 4p\right) }(w|\tau )\text{,}  \label{pro}
\end{equation}
for $l\in \left\{ 0,.,p-1\right\} $. Finally, the operator content of $%
\Gamma _{\theta }$-RCFT $\widehat{u(1)}_{p}$ does not coincide with that of
the ordinary RCFT $\widehat{u(1)}_{4q}$. Indeed, the h.w. representations
corresponding to $a/\sqrt{4q}$ and $\left( a+2q\right) /\sqrt{4q}$, for $%
a\in \left\{ 0,.,q-1\right\} $, of $\widehat{u(1)}_{4q}$ do not belong to $%
\widehat{u(1)}_{q}$.

\end{document}